**Brownian cluster dynamics with short range patchy interactions. Its application to polymers and step-growth polymerization.**

A. Prabhu[1], S. B. Babu[2], J. S. Dolado[3] and J.-C. Gimel[1,1]

*[1] LUNAM Université, Université du Maine, IMMM–UMR CNRS 6283, Département Polymères Colloïdes Interfaces, av. O. Messiaen, 72085 Le Mans Cedex 9, France*

*[2] Malaviya National Institute of Technology Jaipur, Department of Physics, Jaipur, Rajasthan, India*

*[3] Tecnalia Research and Innovation, Derio, Spain*

We present a novel simulation technique derived from Brownian cluster dynamics used so far to study the isotropic colloidal aggregation. It now implements the classical Kern-Frenkel potential to describe patchy interactions between particles. This technique gives access to static properties, dynamics and kinetics of the system, even far from the equilibrium. Particle thermal motions are modeled using billions of independent small random translations and rotations, constrained by the excluded volume and the connectivity. This algorithm, applied to a single polymer chain leads to correct static and dynamic properties, in the framework where hydrodynamic interactions are ignored. By varying patch angles, various chain flexibilities can be obtained. We have used this new algorithm to model step-growth polymerization under various solvent qualities. The polymerization reaction is modeled by an irreversible aggregation between patches while an isotropic finite square-well potential is superimposed to mimic the solvent quality. In bad solvent conditions, a competition between a phase separation (due to the isotropic interaction) and polymerization (due to patches) occurs. Surprisingly, an arrested network with a very peculiar structure appears. It is made of strands and nodes. Strands gather few stretched chains that dip into entangled globular nodes. These nodes act as reticulation points between the strands. The system is kinetically driven and we observe a trapped arrested structure. That demonstrates one of the strengths of this new simulation technique. It can give valuable insights about mechanisms that could be involved in the formation of stranded gels.

## I. INTRODUCTION

The structure and dynamics of a wide range of complex liquids is determined by the aggregation of small particles in solution such as colloids [1-4], proteins [5-7], micelles [8, 9] or oil droplets [10]. Depending on the concentration, the range and strength of the attraction, stable cluster dispersions, transient gels, glasses, phase separated systems can be formed (see for example recent reviews [11, 12]). In order to better understand these processes computer simulations have been done on model systems. They are important tools in colloidal physics giving relevant insights about the structure and dynamical properties, even far from the equilibrium.

---

[1] Author to whom correspondence should be addressed: jean-christophe.gimel@univ-angers.fr

Systems made of attractive particles in solution can be considered from two different viewpoints. Either we speak of aggregates, which form or breakup continuously, or we speak of concentration fluctuations of individual particles. Which of the two approaches appears most natural depends on the system, e.g., in the case of short range attractions one might tend to speak of reversible aggregation, while in the case of long range attractions it may appear more appropriate to speak of concentration fluctuations. Of course, both approaches are strictly equivalent and it is a matter of semantics whether we consider two neighboring particles whose movement is temporarily correlated as belonging to a single transient cluster or as two different particles whose position is correlated by the influence of an attractive potential.

Classical simulations are based on the particle viewpoint. Newtonian dynamic simulations are numerically solving the equations of motion for a set of particles. This algorithm was first proposed for a simple hard sphere fluid in 1957 [13]. It was then extended to other potential like square-well and Lennard-Jones. Most of the time the solvent is ignored and particles have linear trajectories between collisions. Taking the solvent into consideration by representing each of its molecules would be far too expensive in computation time and memory. There are often billions of solvent molecules per particle. To mimic the particle Brownian motion resulting from the random collisions with the solvent molecules, a random force and a friction term are introduced in the equations of motion; leading to either dissipative particle dynamics (DPD) if hydrodynamic interactions are taken into account or Brownian dynamics (BD) if not (see for example [14, 15]). In the case of DPD the friction and the random force felt by a given particle depend on the position of the others. Nowadays BD techniques can deal with a set of $10^4$ particles integrating the equations of the motion over a physical time up to few seconds for micrometric colloidal particles in water at 20°C. This approach is very useful to investigate systems that rapidly reach their equilibrium. Static and dynamical quantities can be computed. For colloids, this is generally the case in the one phase domain of the phase diagram (i.e. for weak attractions, low volume fractions of colloids). But for stronger attractions, when phase separation occurs, the time needed to reach the thermal equilibrium is out of the accessible range using molecular dynamic approaches. In that case Monte-Carlo (MC) simulations based on the Metropolis algorithm [16] are employed to determine the equilibrium state in particular to determine the phase coexistence line [17, 18].

In the past years, we have developed a novel simulation technique called Brownian Cluster Dynamics (BCD), based on the cluster approach. It can handle up to $10^6$ particles during several hours of physical time for micrometric colloidal particles in water at 20°C. It is based on the algorithm originally developed in 1983 by Meakin [19] and Kolb et al. [20] to mimic the irreversible aggregation of Brownian spherical particles which leads to the formation of fractal aggregates. We have extended it to reversible association by allowing already built clusters to break and reform [21, 22]. The algorithm does not involve any resolution of the equations of motion and is only based on a probabilistic approach (see below). It consists of chaining a huge number of times the three following steps: (i) clusters are randomly formed among interacting pairs of particles; (ii) particles undergo random small translational displacements maintaining cluster integrity. This step is very similar to an off-lattice version of the bond fluctuation model [23]; (iii) time is incremented. The algorithm has been described in details elsewhere [24] but will be recalled in the next section to include the patchy interaction. It gives same dynamics as BD simulations and predicts same phase diagrams and static properties as MC simulations [24-26]. Hydrodynamic interactions are not taken into account and hence dynamics is obtained in the free-draining case also called the Rouse limit [27].

Hard sphere fluids interacting via an isotropic short range potential are part of self-assembled materials but they often lead to polydisperse and irregular structures. More precise control over their association to produce predictable structures at the length scale of several particles is an important challenge for novel materials. In comparison, most biological particles assemble into highly monodisperse and precise structures due to the presence of localized and specific interactions at their surface. Following a "biomimetic" approach, an idea was to model those interactions by decorating the surface of the particles with various sticky patches [28] conferring them an anisotropic potential. Nowadays, patchy particles are the subject of growing interest in the self assembled material community and various ingenious particle synthesis techniques have been developed (see for example the review by Pawar and Kretzschmar [29] and recent articles [30-32]). Concomitantly, numerical models have been tested and the influence of that kind of anisotropy on the system behavior has been studied (see for example [28, 33-40]). Sciortino's group has studied the reversible self assembly of particles with two monovalent patches into polymer chains [41, 42]. Their simulation results were remarkably well described by the Wertheim theory [43]. Very recently Marshall and Chapman also developed a new theoretical model to describe self assembling mixtures of single [44] or double [45] patch colloids with spherically symmetric ones.

Here, we would like to present an extended version of the BCD algorithm which takes into account patchy interactions localized on the surface of the particle. In addition to the translational Brownian motion, we have to mimic the rotational one in order to relax both particle positions and patch orientations. The potential we use is similar to a square-well limited to some angular sectors introduced for the first time by Jackson et al. [46] and later studied by Kern and Frenkel [33] and Sciortino's group [36, 42, 47] using computer simulations. To interact, pairs of particles have to be both in range and correctly oriented. Step (ii) of BCD is modified to take into account random small rotations of particles in addition to translational displacements while still keeping bond lengths and orientations in their tolerance domain (cluster integrity). In near future we aim to employ this new algorithm, hereafter denoted patchy Brownian cluster dynamics (PBCD), to study many relevant experimental systems combining strong or irreversible patchy attractions with a superimposed weak attractive isotropic potential (colloidal aggregation of cementitious materials [48], formation of polymeric proteins like gluten [49] or spider silk [50, 51], aggregation of Janus particles [52, 53]). In the present paper we centre our attention in linear polymer chains. On the one hand, the linear architecture of polymers represents one of the most stringent scenarios for illustrating the ability of our computational scheme to capture the underlying physics of anisotropic clusters. On the other hand polymers provide a particularly convenient case of study because many theoretical predictions, simulations and experiments have been made on the subject.

Firstly, using our model, we will revisit static and dynamic properties of linear chains with or without excluded volume effects. In particular we will illustrate how the effects of the chain flexibility (or the lack thereof, stiffness) can be tuned in the algorithm by playing with the cone angle of the patchy interactions, and how the PBCD model can be employed to get insights on the equilibrium configuration of semi-flexible polymers, along with their collective relaxation properties.

Next we will apply the PBCD algorithm to describe the step-growth polymerization; i.e. a type of chemical reactions in which bi-functional monomers react, irreversibly, to form first dimers, then trimers and so on… and eventually long polymer chains. As many naturally occurring and synthetic polymers are produced by step-growth polymerization, e.g. polyesters, polyamides, polyurethanes, etc [54], it is clear that the physics involved in these process is of great practical use. Most of the existing theoretical studies rely on population balance models [55, 56] which keep track of the concentration of every chain length from the knowledge of the starting populations and the effective constant rates [57-59]. In practice, some difficulties appear when assigning values to the effective step-growth chemical rates, as it is still not fully understood how they are influenced by chain-length, diffusion effects and ring formation. In this scenario it is clear that the herein presented PBCD algorithm can serve as a valuable tool for getting a complementary insight, since this new computational scheme naturally account for these mentioned effects. Let us note that in terms of the PBCD algorithm the step-polymerization is merely the irreversible aggregation process of interacting particles with two accessible monovalent patches. Tuning the strength of an additional isotropic potential makes it possible to mimic polymerization in solvent with various thermodynamic qualities. Here, we will present some preliminary results in that sense where polymerization and phase separation are in competition.

The remainder of this paper is organized as follows: Section II describes computational details and applied approximations. Section III presents the results obtained on single polymer chains. Section IV gives some preliminary results about the influence of local flexibility and solvent quality on step-growth polymerization. We will particularly focus on the formation of out of equilibrium arrest networks, very similar to stranded gels obtained with biopolymers (see for example the very recent review by Nicolai and Durand [60] and references therein). Section V summarizes the main findings and gives some conclusions.

## II- MODEL AND SIMULATION TECHNIQUES

### A General case

The patchy potential is modeled using a square-well (SW) potential restricted to some given orientations between both particles [46]. Particles *i* and *j* are in interaction if their distance, $r_{i,j}$, is within the range and the vector, $\mathbf{r}_{i,j}$, linking both particle centers, crosses both patch surfaces. Assuming *d* is the particle diameter, $\varepsilon$ is the relative range of the SW potential and a patch is delimited by a cone of axis $\mathbf{v}_i$ and half opening angle $\omega$ (see figure 1) then the potential, $V(\mathbf{r}_{i,j}, \mathbf{v}_i, \mathbf{v}_j)$, between both particles is given by:

$$V(\mathbf{r}_{i,j}, \mathbf{v}_i, \mathbf{v}_j) = V_1(r_{i,j}) \cdot V_2(\hat{\mathbf{r}}_{i,j}, \hat{\mathbf{v}}_i, \hat{\mathbf{v}}_j),\tag{1}$$

where $\hat{\mathbf{r}}_{i,j} = \mathbf{r}_{i,j}/r_{i,j}$, $\hat{\mathbf{v}}_i = \mathbf{v}_i/v_i$ are unit vectors, $V_1$ is an isotropic SW potential of relative width $\varepsilon$ and depth $u_0 < 0$:

$$V_1(r_{i,j}) = \begin{cases} \infty & r_{i,j} \leq d \\ u_0 & d < r_{i,j} \leq d \cdot (1+\varepsilon) \\ 0 & r_{i,j} > d \cdot (1+\varepsilon) \end{cases},\tag{2}$$

and $V_2$ given by:

$$V_2(\hat{\mathbf{r}}_{i,j}, \hat{\mathbf{v}}_i, \hat{\mathbf{v}}_j) = \begin{cases} 1 & \text{if } \hat{\mathbf{r}}_{i,j} \cdot \hat{\mathbf{v}}_i > \cos\omega \text{ and } -\hat{\mathbf{r}}_{i,j} \cdot \hat{\mathbf{v}}_j > \cos\omega \\ 0 & \text{else} \end{cases}.\tag{3}$$

For a given patch, $\gamma_i$ is the angle between $\mathbf{r}_{i,j}$ and $\mathbf{v}_i$.

The simulation starts at simulation time $t_{sim} = 0$ with an ensemble of *N* randomly distributed and oriented spheres in a box of size $L_{box}$ with periodic boundary conditions. The occupied volume fraction is: $\phi = N \cdot \pi/6 \cdot d^3/L_{box}^3$. A simulation step is made of the three following procedures that are repeated till we reach the desired physical time.

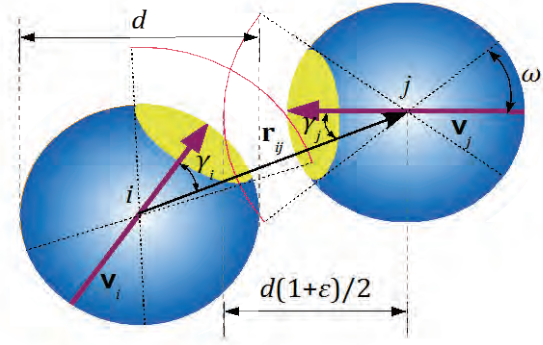

FIG. 1. Representation of the patchy interaction between particles *i* and *j* with diameter *d*. To be in interaction with potential $-u_0$, $\mathbf{r}_{i,j}$ has to cross both patch surfaces (with solid angle $2\pi \cdot (1-\cos\omega)$) and $d < r_{i,j} < d \cdot (1+\varepsilon)$. $\mathbf{v}_i$ defines the patch orientation and $\gamma_i$ is the angle between $\mathbf{r}_{i,j}$ and $\mathbf{v}_i$.

**1. Cluster formation procedure**

Spheres, also named monomers, are considered to be in contact when they are in interaction condition (with correct range and orientation), we call $N_c$ the number of such contacts. In the so-called cluster formation step, monomers in contact are bound with probability $P$. Alternatively, bonds are formed with probability $\alpha$ and broken with probability $\beta$, so that the $P = \alpha/(\alpha+\beta)$. In the latter case, one can vary the kinetics of the aggregation from diffusion limited ($\alpha = 1$) to reaction limited ($\alpha \to 0$) with the same $P$ and thus the same degree of reversibility. Clusters are defined as collections of bound monomers, and individual monomers are clusters of size 1. The ratio of the number of bound ($N_b$) to free contacts ($N_c - N_b$) is given by the Boltzmann factor: $N_b/(N_c - N_b) = \exp(-\Delta E/(k_B \cdot T))$, where $\Delta E$ is the energy difference between a bound and a free contact. The formation of $N_b$ randomly distributed bonds over $N_c$ contacts leads to a decrease in the free energy equal to $-u_0$ per contact (whether bound or free). This decrease may be written as the sum of the decrease in the enthalpy ($\Delta E$) and the gain of the entropy ($\Delta S$): $N_c \cdot u_0 = N_b \cdot \Delta E - T \cdot \Delta S$. The latter is determined by the number of ways $N_b$ bonds can be distributed over $N_c$ contacts: $T \cdot \Delta S = k_B \cdot T \cdot \ln(N_b!/(N_c! \cdot (N_c - N_b)!))$. Noticing that $P = N_b/N_c$, we can express $P$ in terms of $u_0$ [22]:

$$P = 1 - \exp\left(\frac{u_0}{k_B \cdot T}\right) \quad (4)$$

**2. Movement procedure**

The cluster construction procedure is followed by a movement procedure, where $2 \cdot N$ times a sphere is randomly selected and an attempt is made to move it either by an elementary random rotation or translation (with 50% probability each). After $2 \cdot N$ trials, each sphere has been in average entitled to a rotation step and a translation step in an uncorrelated manner. The translational step consists in moving the centre of mass of the sphere in a random direction by a small vector, $\mathbf{s}_T$, where the magnitude of the vector is given by $s_T$. The movement is accepted if it does not lead to an overlap with any other sphere and if it does not break a bond. To achieve the elementary rotation of a given sphere, the tip of its orientation vector $\mathbf{v}$ is performing a random walk on the surface of a sphere of radius $v$. The intensity of the small displacement at the surface is called $s_R$. The movement is accepted only if it does not break any bond or lead to an overlap. It is important to choose step sizes $s_T$ and $s_R$ sufficiently small so the motion is Brownian over the relevant length scales (in the SW and within the cone). To save some CPU time $\mathbf{v}$ can only occupy discrete positions on the surface of the sphere of radius $v$ (see appendix A). The effect of both step sizes on the simulation will be discussed in detail in Sec 2.3.

**3. Physical Time**

After a cluster construction and movement procedure the simulation time ($t_{sim}$) is incremented by one. Calling $t$ the physical time, relationships between $t$, $t_{sim}$, $s_T$ and $s_R$ are obtained considering a free diffusing sphere with the following translational ($D_1^T$) and rotational ($D_1^R$) diffusion coefficients [61]:

$$D_1^T = \frac{k_B \cdot T}{3\pi \cdot \eta \cdot d} \quad (5)$$

$$D_1^R = \frac{k_B \cdot T}{\pi \cdot \eta \cdot d^3} \quad (6)$$

with $k_B$ the Boltzmann constant, $T$ the absolute temperature and $\eta$ the solvent viscosity. For a free particle mean square displacement (MSD), $\langle \mathbf{R}^2(t) \rangle$, is given by [24]:

$$\langle \mathbf{R}^2(t) \rangle = 6 \cdot D_1^T \cdot t = t_{sim} \cdot s_T^2 \quad (7)$$

At short time ($D_1^R \cdot t \ll 1$), the orientation vector **v** performs a two-dimensional random walk on the surface of the sphere and we have [61]:

$$\left\langle \left(\mathbf{v}(t) - \mathbf{v}(0)\right)^2 \right\rangle = v^2 \cdot 4 \cdot D_1^R \cdot t = t_{sim} \cdot s_R^2 \quad (8)$$

The characteristic time, $t_0$, is defined as the time needed for the centre of mass of a free sphere to diffuse a distance equal to its squared diameter: $t_0 = d^2/(6 \cdot D_1^T)$. Using $t_0$, equation (7) gives the following relationship between the physical time, $t$, and the simulation time, $t_{sim}$:

$$\left\langle \mathbf{R}^2(t) \right\rangle / d^2 = t/t_0 = t_{sim} \cdot (s_T/d)^2 \quad (9)$$

The rotational diffusion coefficient of a sphere is simply given by:

$$D_1^R = \frac{1}{2 \cdot t_0} \quad (10)$$

Taking $v = d$ and combining equations (5), (6), (7) and (8) we obtain a simple relationship between $s_T$ and $s_R$:

$$2 \cdot s_T^2 = s_R^2 \quad (11)$$

Equation (9) tells us the simulation time needed to reach a given physical time is inversely proportional to the square of the Brownian step size. Too large steps give too many rejections and slow down artificially dynamics but too small ones restrain the physical time accessible in a reasonable simulation time. A good compromise has to be found between the duration of the simulation and a correct dynamics as we will see in part II.C.

In this simulation study the well-width will be set to 10% of the diameter of a sphere ($\varepsilon = 0.1$) and $\omega$ will be given in radian.

## B. Modeling step-polymerization and polymer chains

As a test case to our model we have done simulation on step-growth polymerization. In this case, two patches are located in opposite directions at the surface of the monomers. As polymerization is a chemical reaction involving covalent bonds the chemical groups form bonds which can be considered irreversible, so $u_0$ is set to infinite and bonds are limited to one per patch. In the cluster construction step, we use $\alpha = 1$ and $\beta = 0$ instead of $P = 1$ to ensure that a previously formed bond is not broken at the expense of the formation of another one when multiple contacts per patch can occur (if $\sin\omega > 0.5/(1+\varepsilon)$). The irreversible reaction thus modeled is equivalent to step-growth polymerization. By varying the half angle of the conic patch ($\omega$), one can control the local rigidity of the polymer chain. Ring formation is also possible which dependents on the local flexibility of the chain.

A polymer chain is made of $m$ hard spherical monomers, indexed by $i$ running from 1 to $m$ (see figure 2). They are linearly and irreversibly linked via SW patches with width $\varepsilon$. The patch orientation of monomer $i$ is characterized by a vector, $\mathbf{v}_i$, pointing toward ascending index along the chain. The bond vector, $\mathbf{r}_i$, ($1 \leq i < m$), connects two consecutive monomers $i$ and $i + 1$ with angles ($\mathbf{r}_i$, $\mathbf{v}_i$) and ($\mathbf{r}_i$, $\mathbf{v}_{i+1}$) smaller than $\omega$.

In order to study in detail static and dynamic properties of a polymer chain, single chains with various $m$ and $\omega$ have been generated in the box. This was realized by randomly choosing positions for the consecutive monomers within bond constraints with or without excluded volume. The next bond vector was chosen with a uniform probability density within the interaction volume (a conic shell). In order to generate unbiased self avoiding cases when excluded volume effects are considered, when a random position leads to an overlap the entire chain is rejected and restarted from the beginning. Obviously this limits the maximum size of chains that can be generated with excluded volume effects, especially when $\omega$ is too high due to an exponential attrition [62].

The measurement of static properties requires only the chain generation step, while dynamics requires many additional movement steps until the chain conformation and position are sufficiently relaxed.

The use of a grid for small patch angles ($\omega < 0.2$ using $q = 80$, see appendix A) leads to some finite size effect biases when generating the next monomer direction. Limiting our choice to only discrete positions makes the occurrence of the exact forward direction (along $\mathbf{v}_i$) too high and increases artificially the amount of bond angles with $\delta = \pi$ exactly. To overcome this problem we had to choose the next direction among a continuous set of values which implied floating point calculations of sines and cosines (while they were pre-calculated on the grid). This method is computationally very expensive and prevents the study of chain dynamics for too small patchs. So static properties were measured down to $\omega = 0.1$ using a continuous set of directions while dynamical properties were measured down to $\omega = 0.2$ using the grid method.

With our patchy polymer chain model (PPC), the bond length of the polymer can fluctuate between $d$ and $d \cdot (1+\varepsilon)$ whose average value is defined by $\langle l_b \rangle = \langle \mathbf{r}_i^2 \rangle^{1/2}$, then the average contour length is given by $L = (m - 1) \cdot \langle l_b \rangle$. Similarly the angle between a patch and a bond can fluctuate between 0 and $\omega$ with its average orientation given by $\langle \cos\gamma \rangle$, and the angle between two consecutive bonds will fluctuate from 0 to $2 \cdot \omega$ with its average value given by $\langle \cos\delta \rangle$. For polymer chains without excluded volume constraints (called ideal chains) all these averages can easily be obtained by considering a uniform distribution of bonds within conic shells. For the completion of this discussion, we are giving the formula of $\langle l_b \rangle$, $\langle \cos\gamma \rangle$ and $\langle \cos\delta \rangle$ in the appendix B. These local averages will be noted with a star in exponent to indicate ideal quantities.

As given in figure 2, we define $G$ as the chain center of mass and $\mathbf{R}_e$ as the end-to-end vector. $\langle \mathbf{R}_{cm}^2(t) \rangle$ is the MSD of the polymer chain center of mass of and $\langle \mathbf{R}^2(t) \rangle$ the MSD of an average monomer in the chain. In this work the relaxation of the chain orientation is followed by monitoring the temporal evolution of the normalized correlation function of the end-to-end vector, defined as:

$$C(t) = \frac{\langle \mathbf{R}_e(t) \cdot \mathbf{R}_e(0) \rangle}{\langle \mathbf{R}_e^2 \rangle} \qquad (12)$$

Monitoring the MSD of the center of mass of the polymer chain gives access to the translational diffusion coefficient of the chain, $D_m^T$:

$$\frac{\langle \mathbf{R}_{cm}^2(t)\rangle}{d^2} = \frac{D_m^T}{D_1^T} \cdot \frac{t}{t_0} \quad (13)$$

In all cases, angular brackets denote averages over all configurations and all possible evolution of the chain. They are obtained by generating at least $10^5$ independent chains for static properties and at least $10^3$ independent configurations and temporal realizations for dynamics.

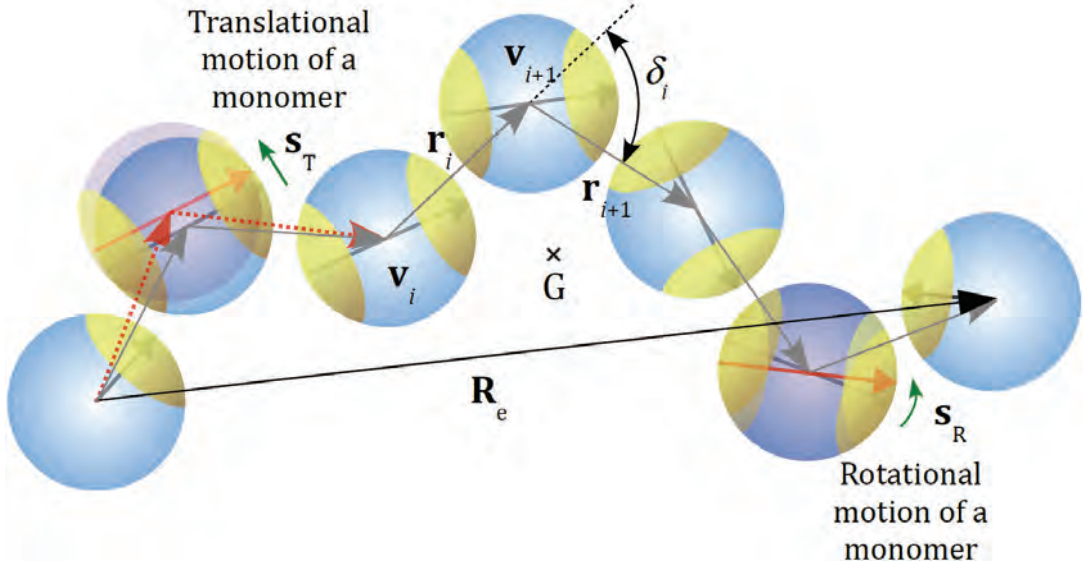

FIG. 2. Schematic representation of a polymer chain using the patchy SW model. $\delta_i$ is the bond angle between two consecutive bonds, $\mathbf{r}_i$ and $\mathbf{r}_{i+1}$, $\mathbf{R}_e$ is the end to end vector and G is the position of the center of mass of the chain. 2 elementary motions are shown on the figure.

In the absence of explicit hydrodynamic interactions, dynamics of long ideal flexible polymer chains is described by the so-called Rouse equation that forms the basis of all analytical calculations [27, 63, 64]. Panja et al. [65] have recently proposed an approximate analytical expression for the mode amplitude correlation functions for long flexible polymer chains with or without exclude volume. This gives for $C(t)$:

$$C(t) = \sum_{p \text{ odd}} A(\tau_p) \cdot \exp(-t/\tau_p) \quad m \gg 1 \quad (14)$$

with

$$\tau_p = \frac{\tau_{max}}{p^{1+2\nu}}, \quad \tau_{max} \propto m^{1+2\nu} \tag{15}$$

and

$$A(\tau_p) = \left(p^{1+2\nu}\right)^{-1} \Big/ \sum_{p\,\text{odd}} \left(p^{1+2\nu}\right)^{-1} \tag{16}$$

with $\nu$ the Flory exponent. $\nu = 1/2$ for ideal chains and a mean field approach gives $\nu \approx 3/5$ for self-avoiding ones. A more sophisticated derivation leads to $\nu = 0.588$ for self avoiding chains (see [66] for more details). $p$ is the mode index and only odd modes contribute to the relaxation of $C(t)$. The first mode ($p = 1$) has the longest relaxation time and is called $\tau_{max}$. The subsequent relaxation modes ($p = 3, 5...$) have smaller intensities and relaxation times by a factor $p^{1+2\nu}$ as written in equations (15) and (16).

TABLE I. Relaxation times and their contribution to $C(t)$ for a long ideal chain and self avoiding one. See text and equations (14), (15) and (16).

|   | $\nu = 0.5$ | | $\nu = 0.588$ | |
|---|---|---|---|---|
| $p$ | $\tau_{max}/\tau_p$ | $A(\tau_p)$ | $\tau_{max}/\tau_p$ | $A(\tau_p)$ |
| 1 | 1 | 0.811 | 1 | 0.853 |
| 3 | 9 | 0.090 | 10.9 | 0.078 |
| 5 | 25 | 0.032 | 33.2 | 0.026 |
| 7 | 49 | 0.017 | 69.0 | 0.012 |
| 9 | 81 | 0.010 | 119.2 | 0.007 |
| 11 | 121 | 0.007 | 184.5 | 0.005 |

Table 1 gives the expected relative contribution of each odd mode, $A(\tau_p)$, together with the ratio $\tau_{max}/\tau_p$ for both values of $\nu$. We clearly see that the main contribution arises from $\tau_{max}$ that is responsible for more than 80% of the relaxation of $C(t)$. The 2nd relaxation time ($p = 3$) is around ten times smaller than $\tau_{max}$ with a contribution less than 10%. This means that for $t > \tau_{max}$, $C(t)$ behave mainly as a single exponential:

$$C(t) \simeq A(\tau_{max}) \cdot \exp(-t/\tau_{max}), \quad t \gg \tau_{max} \tag{17}$$

The rotational diffusion coefficient of a polymer chain, $D_m^R$, is calculated using the longest relaxation time $\tau_{max}$:

$$D_m^R/D_1^R = t_0/\tau_{max} \quad (18)$$

For rigid objects, $C(t)$ relaxes as a single exponential with $\tau_{max} \propto m^3$ [63].

The distribution of relaxation times can be obtained by a regularized Inverse Laplace Transform (ILT) of $C(t)$ using a constrained regularization calculation algorithm called REPES [67]. ILT is notoriously mathematically ill-conditioned [68] and is very sensitive to noise in the data [69]. But considering the expected shape of the distribution (the first mode is the main contribution and is well separated from the third one) this technique is sufficiently accurate to extract a correct value for $\tau_{max}$ if $C(t)$ is relaxed around 1%. Figure 3 shows an example for a polymer chain with excluded volume, $m = 20$, $\omega = 1.0$ and $s_R/d \approx 0.0186$. In order to properly calculate $\tau_{max}$, the simulation has been carried out on at least thousand chains for times up to $t = 4.6 \cdot \tau_{max}$. Considering our computational resources, this could be achieved in a reasonable CPU time for chains with size only up to $m = 20$. From these limitations, our algorithm is clearly not designed to study dynamics of long chains which is not the purpose of our study.

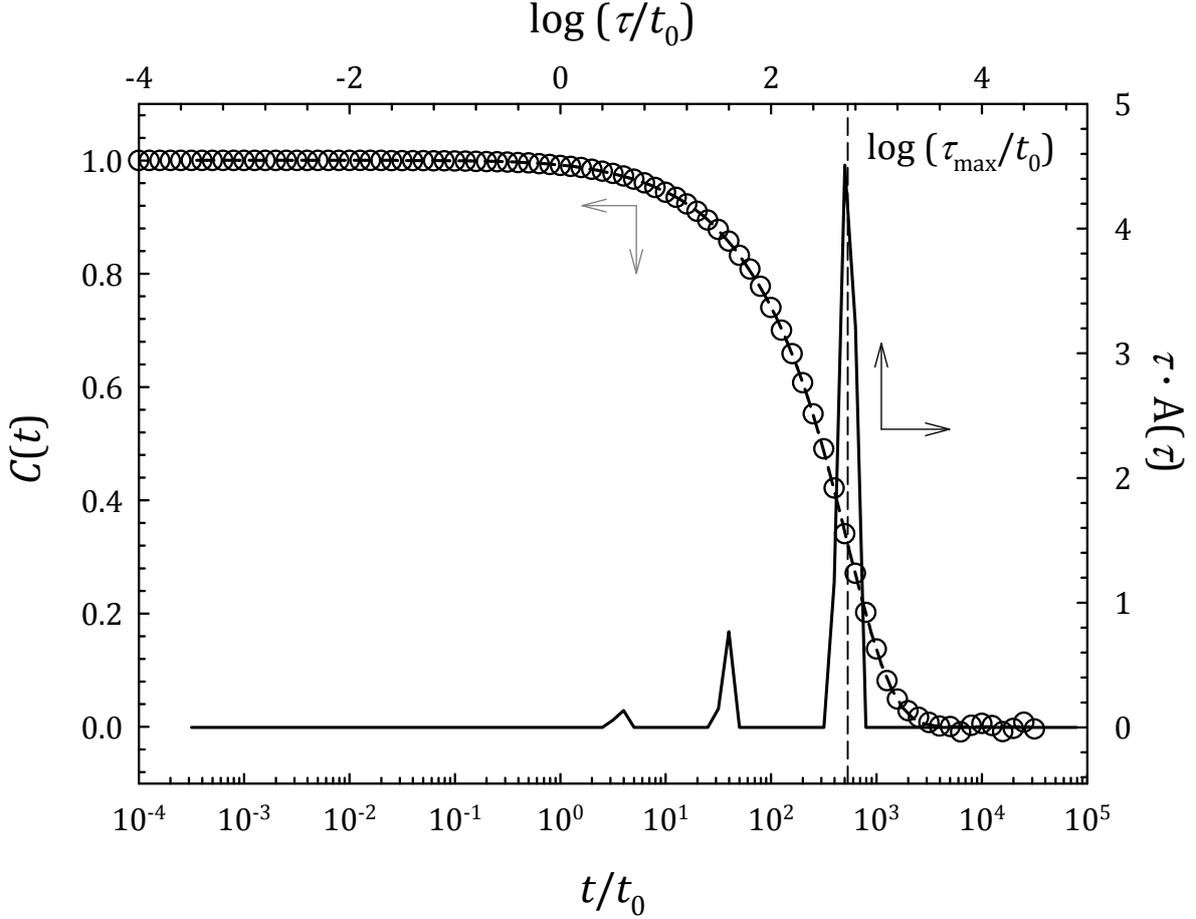

FIG. 3. Representation on a lin-log plot of the end to end vector correlation function, $C(t)$ (open circles, left and down axis), superimposed with the corresponding relaxation time distribution, $\tau \cdot A(\tau)$ obtained by ILT using REPES [67] (solid line, right and top axis). It was measured on a polymer chain with excluded volume, $m = 20$, $\omega = 1.0$ and $q = 65$ ($s_R/d \approx 0.0186$), see appendix A. Error bars on $C(t)$ are of the order of symbol size. The dashed line is the fitted curve obtained from REPES. The vertical dotted line shows $\tau_{max}/t_0 \approx 535$.

## C. Finite size effects

As we discussed previously, the Brownian step size, $s_R$ or $s_T$ (both being related by equation (11)) has to be as small as possible to give correct dynamics. But a too small value of the step size slows down the calculation by increasing the number of simulation steps needed to reach the same physical time (see equation (9)). In figure 4(a), we have plotted the translational diffusion coefficient of the center of mass of a small polymer chain ($m = 7$ with excluded volume interactions) as a function of $s_T$. We clearly see that as $s_T$ increases the translational diffusion coefficient is reduced. Monomers that belong to the chain have to respect two constraints in order to move. They should not break any bond nor should overlap. If such an event occurs their motion is rejected but the simulation time is still incremented. As the step size increases, monomers face more and more potential rejections and dynamics is artificially slowed down. In the range of step sizes used, we observe a linear dependency of this phenomenon and that the diffusion coefficient converges toward a finite value at $s_T = 0$, which we define to be the real translational diffusion coefficient of the chain. A similar behavior is observed for rotational diffusion (see figure 4(b)). In the following, all results given for dynamics of a single chain have been extrapolated to zero step size. But using such an extrapolation to study the complete polymerization reaction would be computationally too expensive and we have to define what we consider to be an acceptable step size. We arbitrarily state that to be acceptable apparent dynamics should be within 10% of the real one. We already had to face that problem before in the case of BCD, but with isotropic potential and no rotational motion. We found that to be acceptable, the translational step size had to be small enough compared to any characteristic length scale in the system (see [25, 70] for details). This means the step size has to be small compared to the well width but also compared to the average distance to the first neighbor ($\Delta$) when the simulation starts with a random distribution of hard spheres. Introducing the rotational motion adds another characteristic length that is the size of the patch which delimits the accessible positions to the tip of the orientation vector. Combining all these criteria with the constraint to be within 10% of real dynamics leads to the following condition: $s_R/d < \omega/10$ and $s_T/d < \varepsilon/5$ and $s_T < (\Delta-1)/3$.

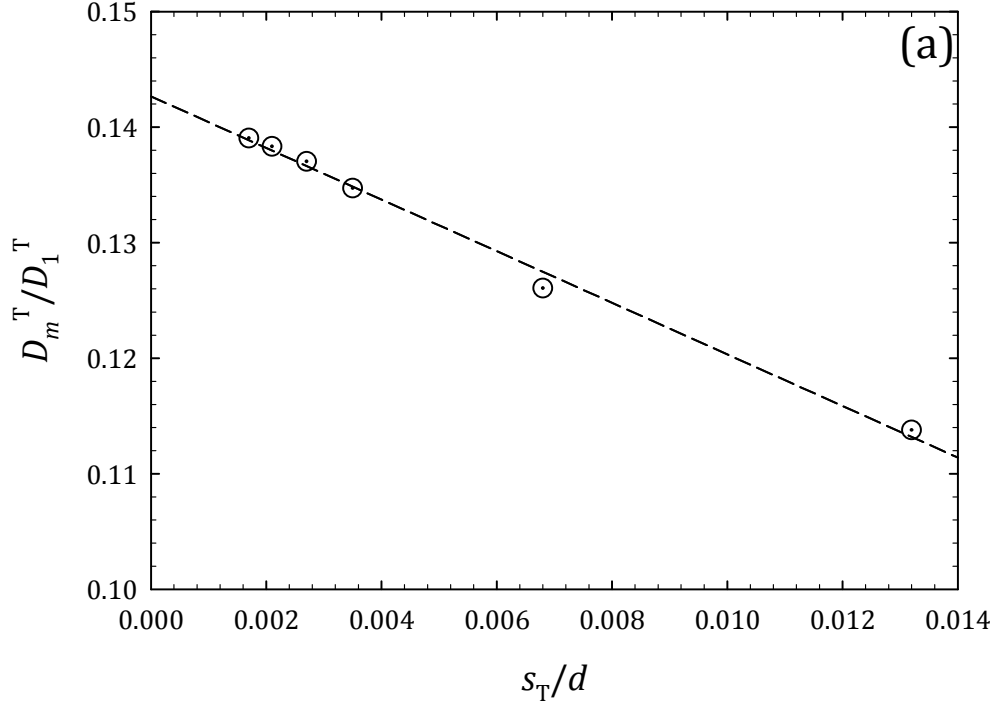

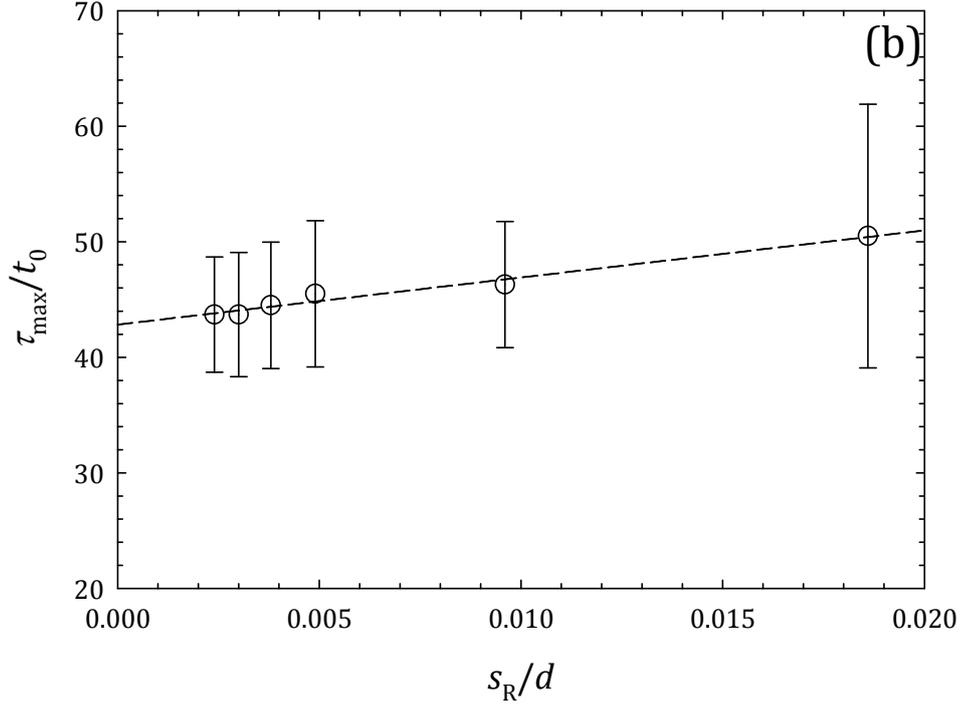

FIG. 4. (a) Evolution of the translational diffusion coefficient of a polymer chain ($m = 7$, $\omega = 0.2$) as a function of the translational Brownian step size, $s_T$. $D_m^T/D_1^T$ has been calculated using equation (13) from the MSD of the chain center of mass. Error bars are of the order of symbol size. (b) Evolution of the longest relaxation time of a polymer chain ($m = 5$, $\omega = 0.2$) as a function of the rotational Brownian step size, $s_R$. $\tau_{max}$ is obtain from the ILT of the end to end vector correlation function giving a relative error from 10% to 20% on the measurement of $\tau_{max}$.

## III. STATIC AND DYNAMIC PROPERTIES OF A SINGLE POLYMER CHAIN USING THE PATCHY MODEL

### A. Influence of excluded volume effects on the local structure of the chain

As expected, local quantities obtained for generated ideal chains are found to be exactly the one predicted by equations (33), (34), and (35) in appendix B. They also do not depend on the size of the chain. For a given $\omega$, deviations from ideality increase with the size of the chain and stabilize on a plateau value at large $m$ defining an asymptotic behavior. In figure 5, 6 and 7, these asymptotic values have been compared to ideal ones (noted with an asterisk, see appendix B) and plotted as a function of $\omega$. We clearly see that excluded volume effects have a very little influence on the local structure, especially as $\omega \leq 1$, even if it is systematic. $\langle l_b \rangle$ is only very slightly influenced even at large $\omega$ (around 0.04% at maximum for $\omega = \pi$) and will be considered a constant ($\langle l_b \rangle / d = 1.052$). $\langle \cos\delta \rangle$ is much more sensitive on $\omega$ than $\langle \cos\gamma \rangle$ and reaches a limited value $\langle \cos\delta \rangle \approx 0.295$ for freely jointed beads with excluded volume ($\omega = \pi$, see figure 7(a)).

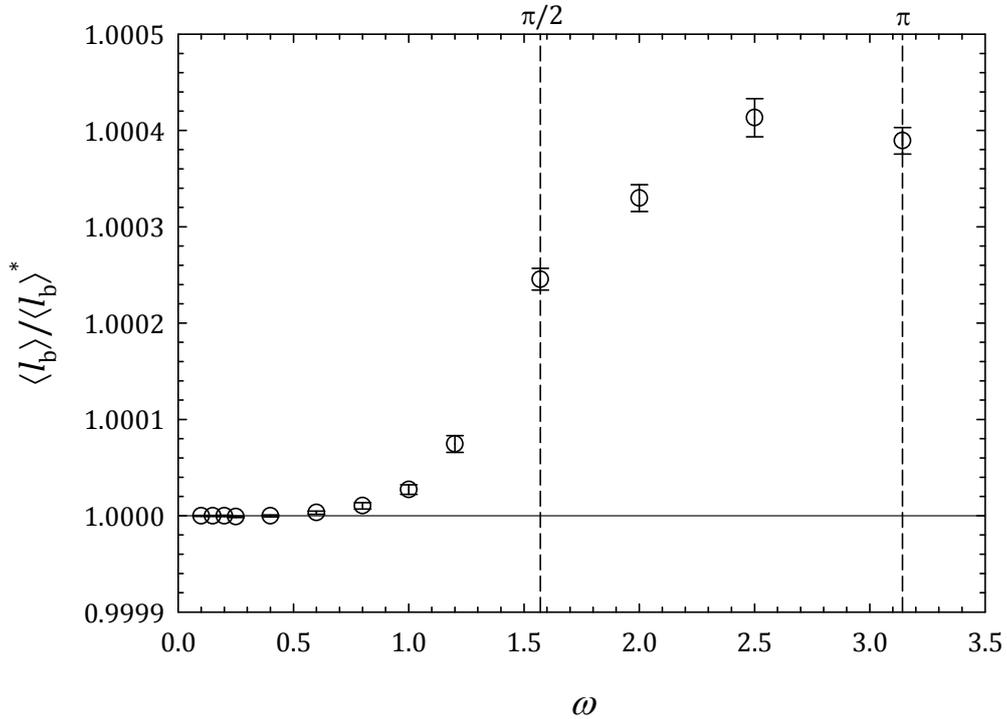

FIG. 5. Evolution of $\langle l_b \rangle / \langle l_b \rangle^*$ as a function of $\omega$ in the limit of large self avoiding chains.

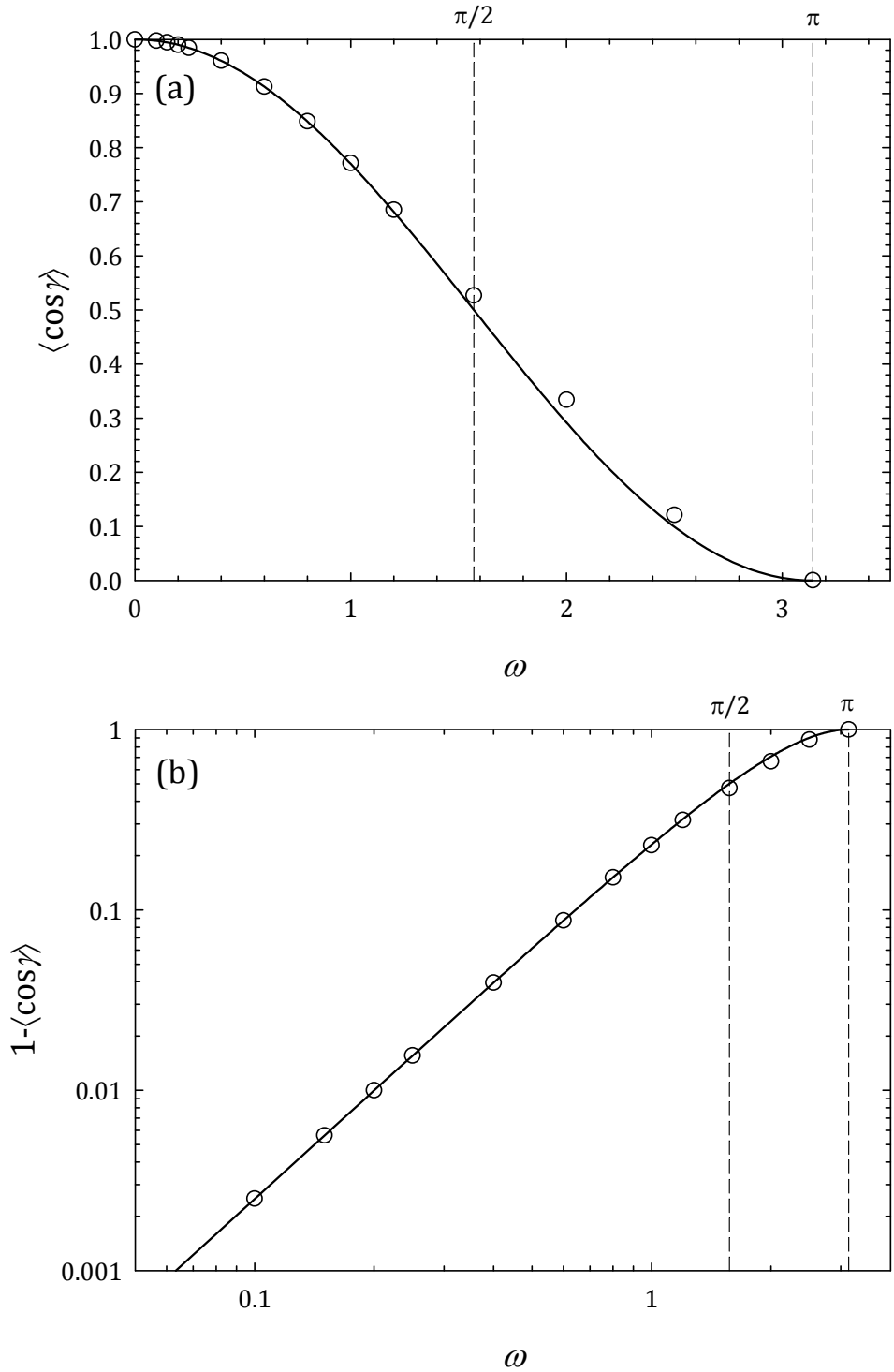

FIG. 6. (a) Evolution of $\langle \cos \gamma \rangle$ as a function of $\omega$ in the limit of large self avoiding chains. (b) Same data but plotting $1-\langle \cos \gamma \rangle$ on a log-log scale. In both cases, the solid line represents ideal predictions from equation (34). Error bars are smaller than symbol size.

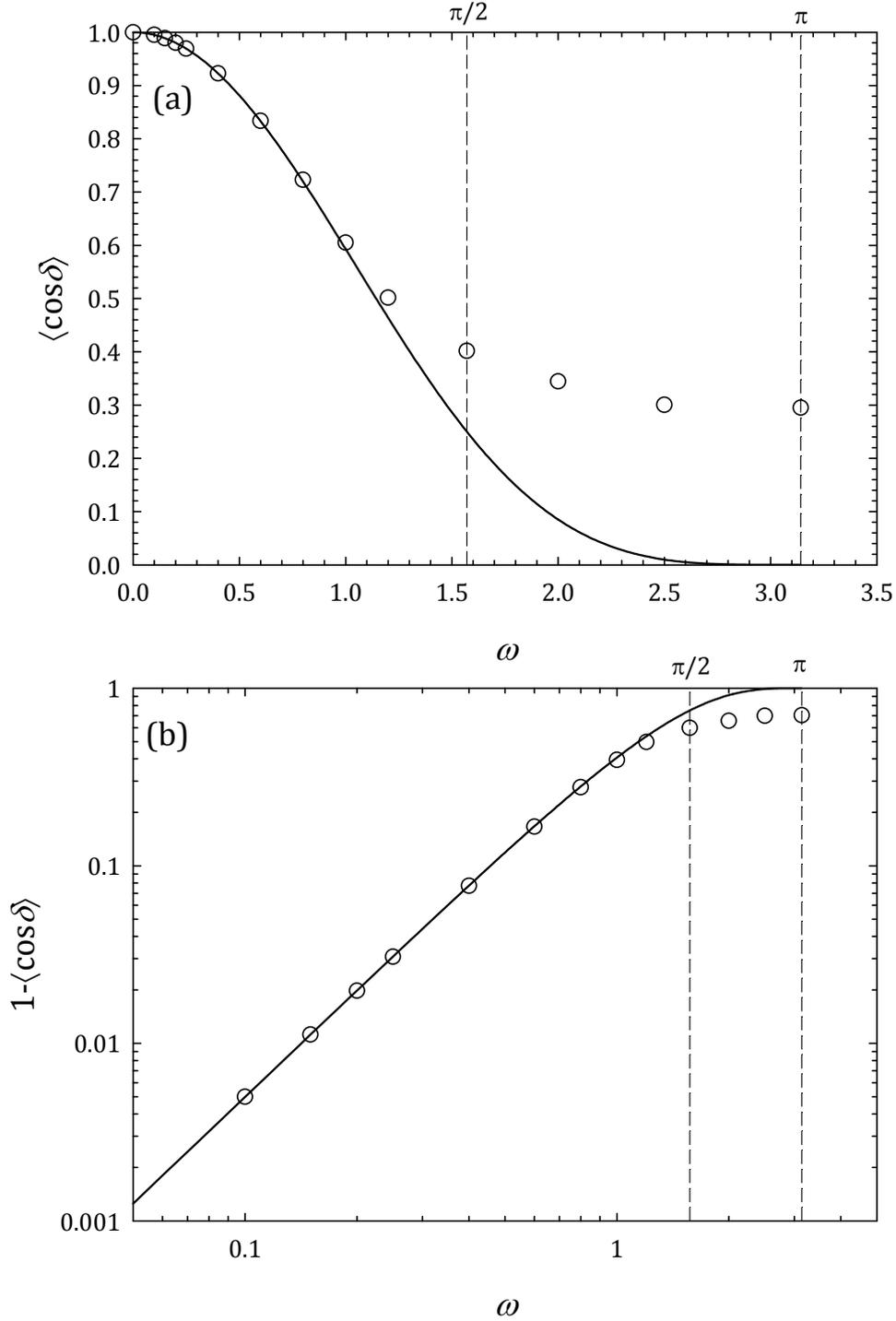

FIG. 7. (a) Evolution of $\langle \cos \delta \rangle$ as a function of $\omega$ in the limit of large self avoiding chains. (b) Same data but plotting $1-\langle \cos \delta \rangle$ on a log-log scale. In both cases, the solid line represents ideal predictions from equation (35). Error bars are smaller than symbol size

## B. The patchy polymer chain and the freely rotating chain model

As our model has well defined average bond lengths and bond angles, the PPC model presents intrinsic analogies with the classical freely rotating chain (FRC) model described by Flory [71], built with a fixed bond length, $l_b$ and a fixed angle, $\delta$. In the limit of very small bond angles the FRC model is called the worm like chain (WLC) model. In that case, the chain is viewed as a continuous line, more or less flexible, with no thickness. The PPC model with $\omega = \pi$ is the analogue of the freely jointed chain (FJC) model. Assuming no excluded volume effects, the average square end to end distance, $\langle \mathbf{R}_e^2 \rangle$, and the square radius of gyration, $\langle R_g^2 \rangle$, can be calculated and a persistent length $l_p$ can also be defined: $l_p/l_b = -1/\ln(\cos\delta)$ (see appendix C). Figure 8 shows that whatever $m$ or $\omega$, results obtained for ideal patchy polymer chains superimpose exactly with predicted value from the FRC model if $\langle \cos\delta \rangle$ is used instead of $\cos\delta$. The ratio, $X = L/l_p$, used in the figure is simply given by $X = -(m-1) \cdot \ln(\langle \cos\delta \rangle)$. An ideal PPC can be viewed as an ideal FRC using $\langle \cos\delta \rangle$ and $\langle l_b \rangle$ instead of fixed quantities.

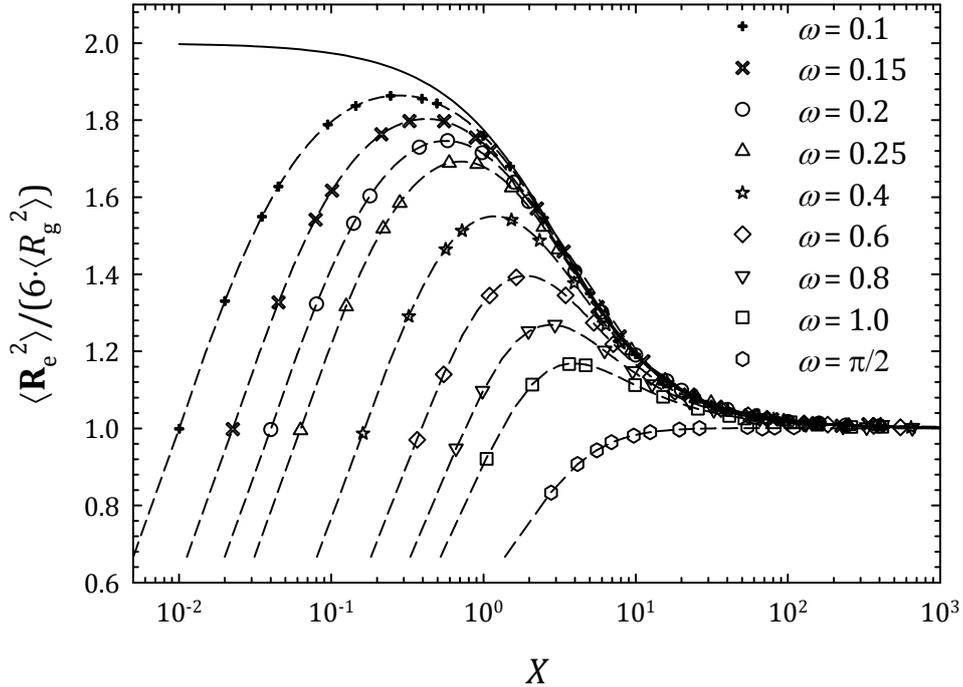

FIG. 8. Evolution of the ratio $\langle \mathbf{R}_e^2 \rangle/(6 \cdot \langle R_g^2 \rangle)$ as function of $X = L/l_p$ for ideal patchy polymer chains with various cone angle as indicated in the figure. $\langle \cos \delta \rangle$ has been used to calculate $l_p$ (see text). The solid line represents the expected behavior for a WLC. Dotted lines represent theoretical predictions from the FRC model (see appendix C).

### C. Influence of excluded volume interactions on the bond correlation function

It is well known that introducing excluded volume interactions lead to the swelling of the chain which leads to increasing characteristic sizes of the polymer chain (end to end distance, radius of gyration…). This effect is not a simple proportionality factor but also changes the Flory exponent involved in the description of the asymptotic scaling of $\langle \mathbf{R}_e^2 \rangle$ with the number of monomer in the chain [72]:

$$\langle \mathbf{R}_e^2 \rangle \propto m^{2\cdot\nu} \quad m \gg 1 \tag{19}$$

Despite average local quantities are very little influenced by that effect when $\omega$ is small, the bond correlation function, $\langle \cos\theta(n) \rangle$ is strongly modified. $\langle \cos\theta(n) \rangle$ is the average cosine of the bond angle between bond $\mathbf{r}_i$ and $\mathbf{r}_{i+n}$. For an ideal chain, it is a single relaxing exponential with a relaxation length $l_p$ given in appendix C. But for large self avoiding chains, $\langle \cos\theta(n) \rangle$ must scale as:

$$\langle \cos\theta(n) \rangle \propto n^{2\cdot\nu - 2} \quad n \gg 1 \tag{20}$$

so the correct scaling from equation (19) is recovered [73].

In figure 9 we have plotted $\langle \cos\theta(n) \rangle$ for a self avoiding chain with $m = 2\cdot 10^4$ and $\omega = 0.2$. We clearly see that up to a given $n$, the relaxation is the same as the one of an ideal chain with a persistent length calculated using equation 35 and 45 (see appendix C):

$$l_p / \langle l_b \rangle = -\left(2\cdot\ln\left(\frac{1+\cos\omega}{2}\right)\right)^{-1} \tag{21}$$

Then a crossover takes place and equation (20) is recovered. For a chain with finite length, as $n$ approaches its maximum value ($n = m-2$) we observe a deviation from the power law and a final cut-off occurs.

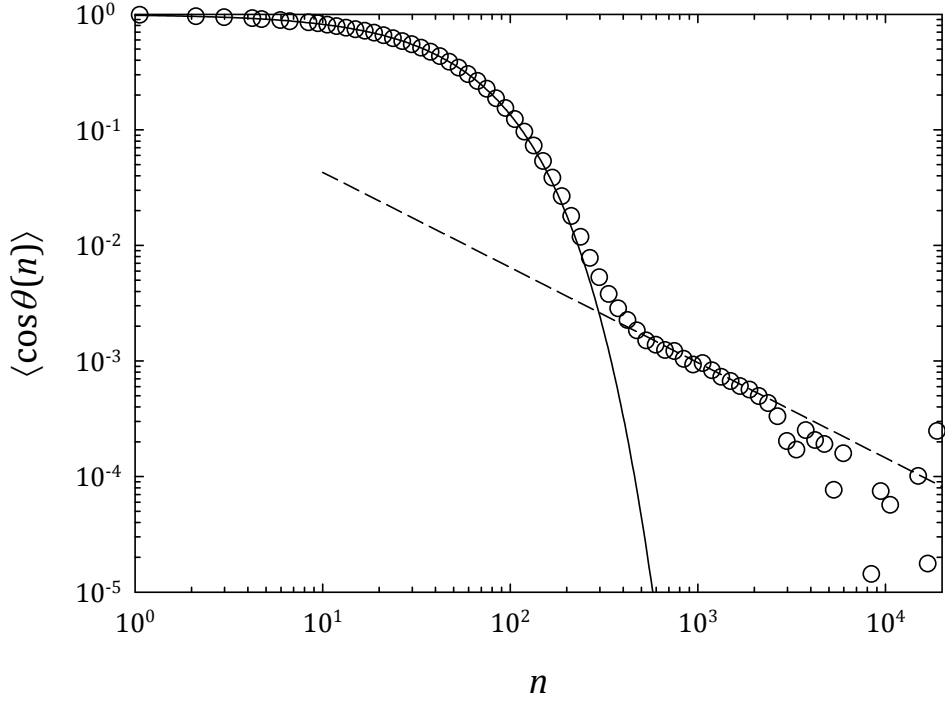

FIG. 9. Evolution of the bond correlation, $\langle\cos\theta(n)\rangle$, as a function of the bond index separation, $n$, for a self avoiding patchy polymer chain with $m = 2\cdot 10^4$ and $\omega = 0.2$. The solid line is the ideal prediction for the corresponding parameters. The dotted line has a slope of $2\cdot\nu-2 = -0.824$.

In figure 10 we see that the complete behavior of the bond correlation can be remarkably adjusted all along using the following phenomenological equation:

$$\langle\cos\theta(n)\rangle = A\cdot\exp(-n/n_1) + (1-A)\cdot(1+n/n_1)^{2\cdot\nu-2}\cdot\exp(-n/n_2) \qquad (22)$$

where $A$, $n_1$ and $n_2$ are fitting parameters. Figure 10 shows the fit made on a self avoiding chain with $m = 8\cdot 10^3$ and $\omega = 0.4$ using a nonlinear regression with the Marquardt-Levenberg algorithm (see for example [74]). For a given $\omega$, values obtained for $n_1$ are independent on $m$ as soon as $m \gg n_1$, see figure 11(a). Figure 11(b) shows that $n_1$ is indeed equal to $l_p/\langle l_b\rangle$ where $l_p$ has been calculated using $\langle\cos\delta\rangle$:

$$l_p/\langle l_b\rangle = -1/\ln(\langle\cos\delta\rangle) \qquad (23)$$

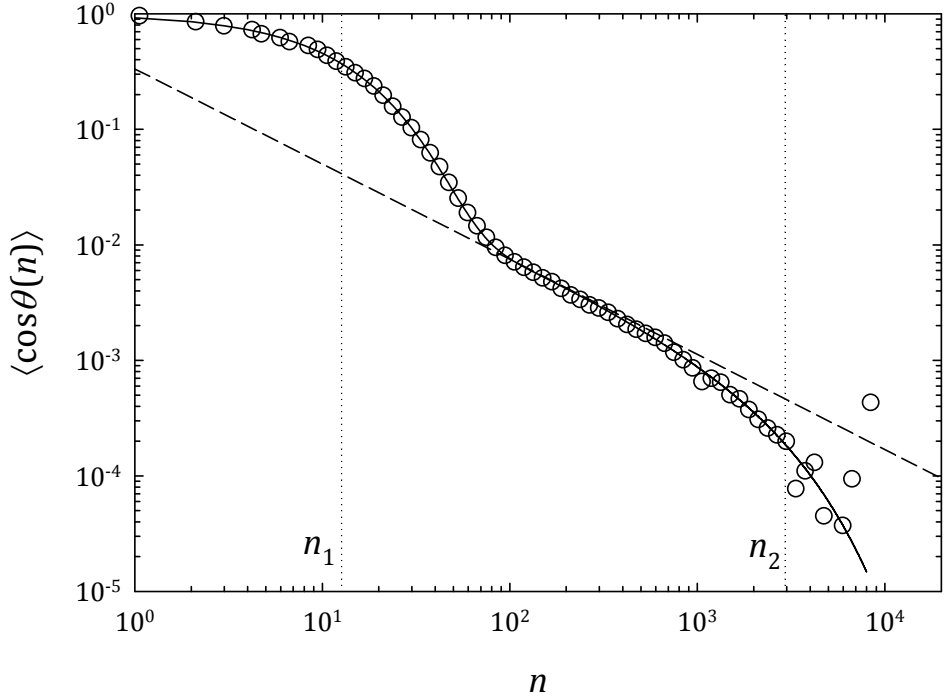

FIG. 10. Evolution of the bond correlation, $\langle \cos\theta(n) \rangle$, as a function of the bond index separation, $n$, for a real patchy polymer chain with $m = 8 \cdot 10^3$ and $\omega = 0.4$. The solid line is a fit to equation (22). The dashed line has a slope of $2 \cdot \nu - 2 = -0.824$. The two vertical dotted lines figure out $n_1$ and $n_2$ values obtained from the fit. In that example we have $A = 0.9537 \pm 0.001$, $n_1 = 12.69 \pm 0.01$ and $n_2 = 2938 \pm 33$.

Note that equations (21) is equivalent to equation (23) as soon as $\omega$ is small enough (see figure 8). For semi-flexible PPC with excluded volume interactions, the exponential relaxation at small $n$, independent on $m$, makes relevant the use of $l_p/\langle l_b \rangle$ to characterize the local intrinsic stiffness of the chain. Other classical persistent lengths, like the Kuhn length or the integral of the bond correlation function diverge with $m$ and cannot be used to characterize the local flexibility for self avoiding chains (see [75] for details). In the following, we have fixed $n_1 = l_p/\langle l_b \rangle$ in the fitting procedure. This leads to more accurate determinations for $A$ and $n_2$. Figure 12(a) shows that $A$ does not depend on $m$, when $m$ is large enough; while figure 12(b) shows that the cut-off value $n_2$ is simply proportional to $m$. The prefactor of the power law in equation (20) is given by $(1-A) \cdot n_1^{2-2\nu}$ in equation (22).

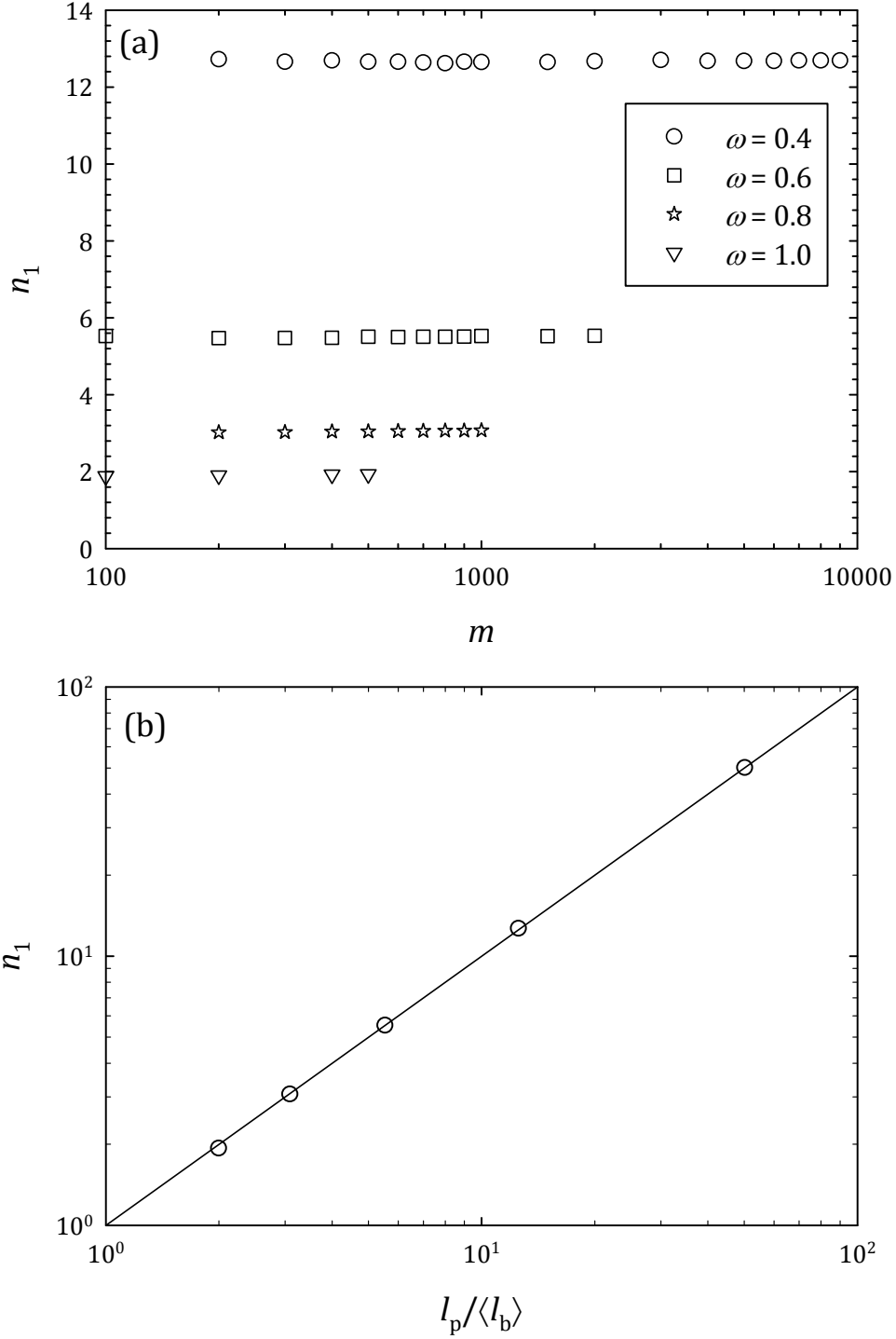

FIG. 11. (a) Evolution of the fitted parameter $n_1$ (see equation (22)) as a function of $m$ for various $\omega$. (b) Evolution of $n_1$ as a function of $l_p/\langle l_b \rangle$ calculated using $\langle \cos\delta \rangle$ (see equation (23)). The solid line figures $x = y$. In both cases, error bars are smaller than the symbol size.

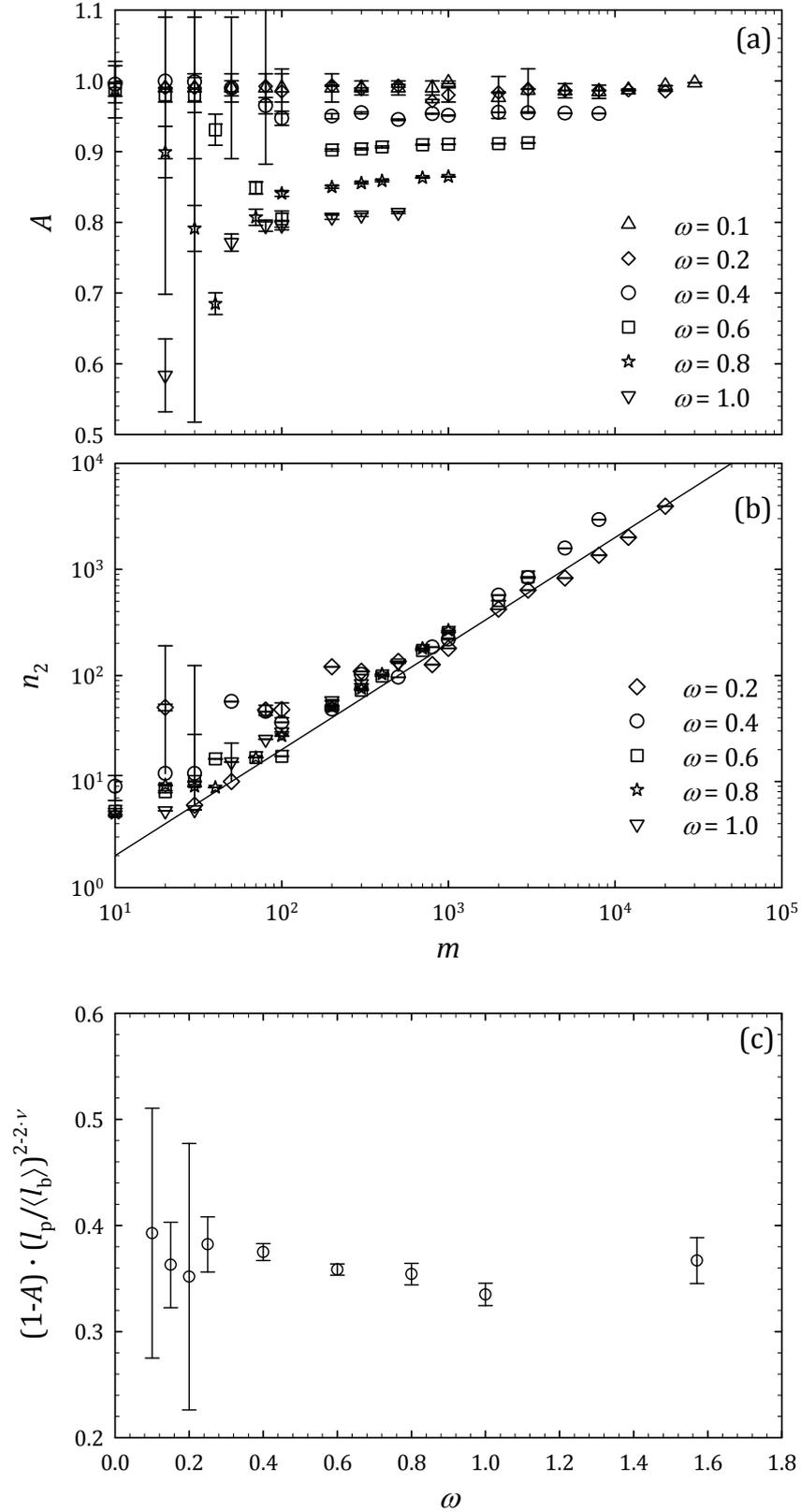

FIG. 12. (a) and (b) Evolution of the fitted parameters $A$ and $n_2$ respectively, using $n_1 = l_p/\langle l_b \rangle$ (see text). The solid line in (b) has a slope of 1. (c) Evolution of the prefactor of the power law in equation (20) as a function of $\omega$.

Figure 12(c) shows it doesn't depend significantly on $\omega$ and stays close to 0.4. A similar behavior was found by Hsu et al. [76] but using a non realistic model of a semi flexible self-avoiding chain on a cubic lattice. In their model, they define a bending energy, $u_b \cdot (1-\cos\delta)$, with $\delta = 0$ or $\delta = \pm\pi/2$ on the cubic lattice. $u_b = 0$ corresponds to the classical self-avoiding walk while increasing $u_b$ leads to the formation of stiffer chains. They also found a primary exponential decay for $\langle\cos\theta(n)\rangle$ with a relaxation index proportional to $\exp(u_b/(k_B \cdot T))$, characteristic of the local rigidity of the chain (see figure 1(a) in [76]). Then $\langle\cos\theta(n)\rangle$ reaches the expected power law behavior predicted by equation (20) but with a different prefactor. Our PPC model can be considered as an off-lattice version of their on-lattice model and the prefactor of equation (20) appears to be dependent on the model used to mimic semi flexible self avoiding chains.

Finally, figure 13 shows $l_p$, calculated with equation (23), as a function of $\omega$ for PPC with excluded volume interactions together with the theoretical prediction for ideal chains. In the range $0.1 \leq \omega \leq \pi$, the persistent length of the self-avoiding PPC model varies over 2 decades. In the limit of small $\omega$, $l_p$ is not influenced anymore by excluded volume effects and scales like:

$$l_p / \langle l_b \rangle \simeq 2/\omega^2 \quad \omega \ll 1 \tag{24}$$

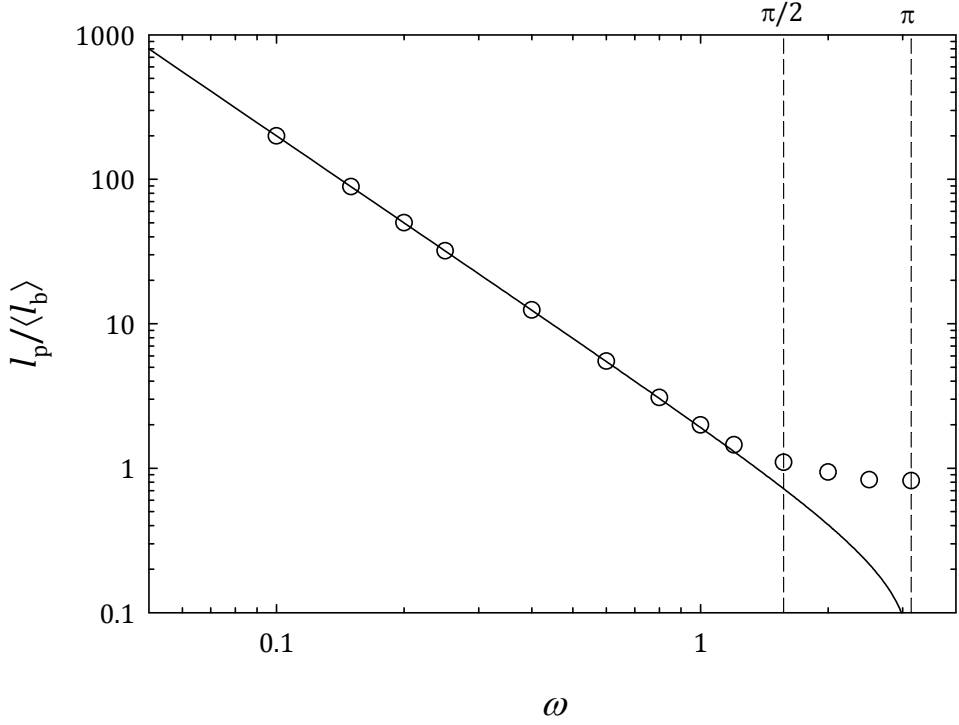

FIG. 13. Evolution of the ratio $l_p/\langle l_b \rangle$ as a function of $\omega$ for excluded volume chains. The solid line represents the predicted ideal behavior (see equation (21)). Error bars are smaller than the symbol size

### D. Deviation from ideality, influence of the local flexibility on the thermal blob size

In order to study the deviation from ideality, $\langle \mathbf{R}_e^2 \rangle$ has been normalized by its ideal value, $\langle \mathbf{R}_e^2 \rangle^*$ (see appendix C), and plotted for various $m$ and $\omega$. Figure 14(a) shows the evolution of that ratio as a function of $L/d$ and figure 14(b) as a function of $X = L/l_p$, with $l_p$ calculated using equation (23) for self avoiding chains. It is clear that, as $\omega$ decreases, self avoiding semi flexible chains present an ideal behavior before a cross-over to swollen chains takes place. This appears at higher values of $X$ as $\omega$ decreases. For the smallest value, $\omega = 0.10$, chains up to $m = 3 \cdot 10^4$ are still insensitive to excluded volume effects and behave as ideal ones even far in the flexible regime ($X \approx 150$). Moreover, $l_p$ is not the relevant length scale that describes the cross-over from ideal to swollen chains.

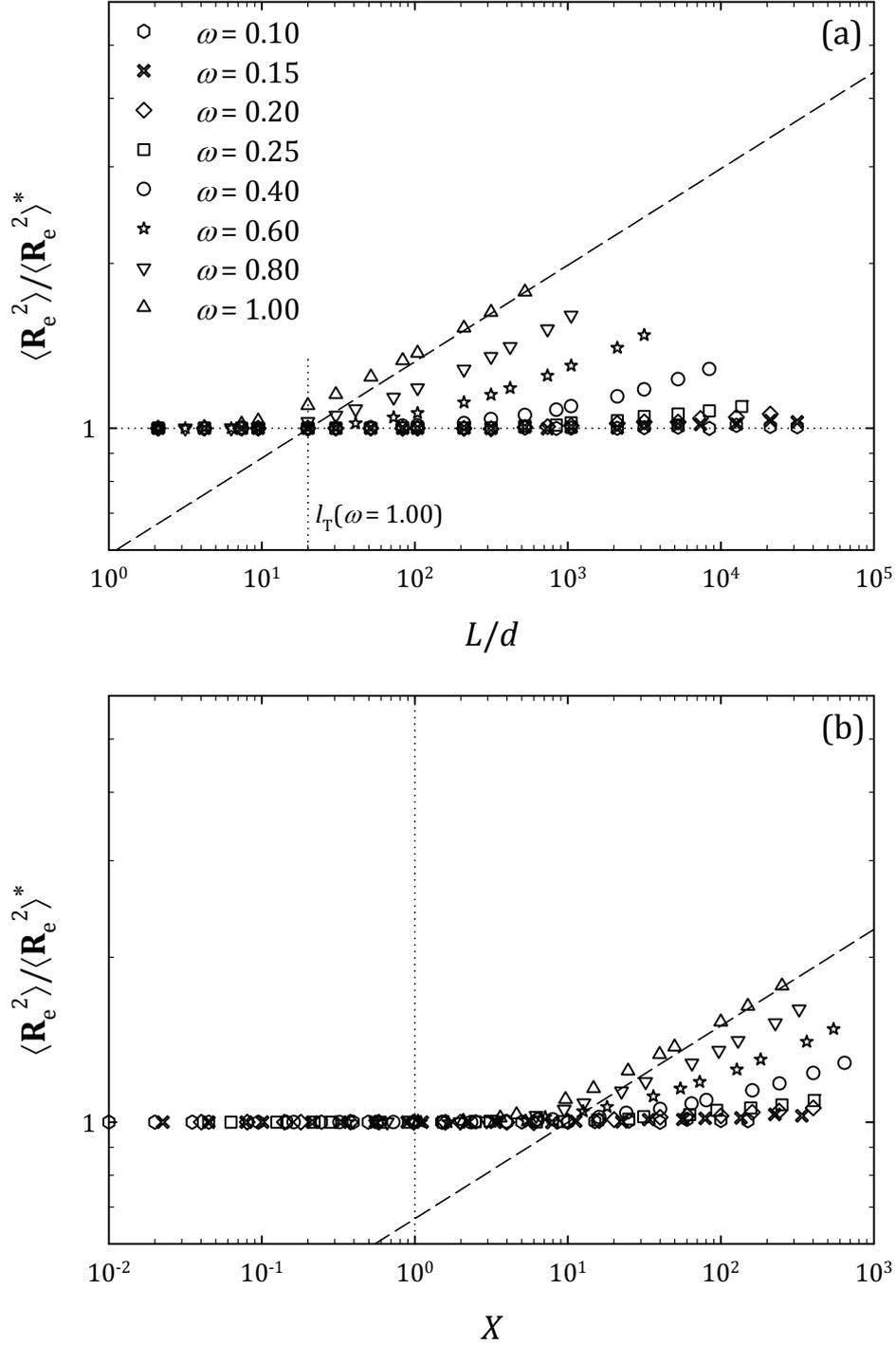

FIG. 14. (a) Log-log representation of the ratio $\langle \mathbf{R}_e^2 \rangle / \langle \mathbf{R}_e^2 \rangle^*$ as a function of $L/d$ for various $\omega$ as indicated in the figure. The dashed line has a slope $2 \cdot \nu - 1 = 0.176$. It shows the limiting behavior of $\langle \mathbf{R}_e^2 \rangle / \langle \mathbf{R}_e^2 \rangle^*$ for $\omega = 1.00$. Dotted lines indicates the intersection of the previous power law with the line $\langle \mathbf{R}_e^2 \rangle / \langle \mathbf{R}_e^2 \rangle^* = 1$. (b) Same data but as a function of $X = L/l_p$. The dotted vertical line figures out the limit between rigid and flexible chains. Error bars are smaller than the symbol size.

Using scaling arguments for semi flexible chains with excluded volume interactions, Schaefer et al. [77] have proposed an explanation to describe the length scale, $l_T$, at which that cross-over takes place. They use the thermal blob concept and assume the chain is made of blobs, with characteristic squared length scale $\langle \xi_T^2 \rangle$, freely jointed in a self avoiding manner. Inside each blob, containing $m_T$ monomers, the chain is considered as ideal with persistent length, $l_p$, and bond length $l_b$. $l_T$ is simply the contour length inside a blob, given by: $l_T = (m_T-1) \cdot l_b$. On one hand, considering that kind of structure leads to the following relationship:

$$\langle \mathbf{R}_e^2 \rangle \propto \left(L/l_T\right)^{2\cdot\nu} \cdot \langle \xi_T^2 \rangle \propto \left(L/l_T\right)^{2\cdot\nu} l_T \cdot l_p \qquad (25)$$

On the other hand, taking a Flory approach to calculate the free energy of the semi flexible chain and minimizing it gives for very long chains in good solvent (high temperature limit see equation (7) in [77]):

$$\langle \mathbf{R}_e^2 \rangle^{1/2} \propto m^{3/5} \cdot \left(l_p/l_b\right)^{1/5} \cdot l_b \qquad (26)$$

Taking $\nu = 3/5$ and combining equations (25) and (26) Schaefer et al. [77] predict:

$$l_T \propto \left(l_p/l_b\right)^3 \cdot l_b \qquad (27)$$

Unfortunately, for the smallest $\omega$, we cannot generate sufficiently large chains such that the swollen behavior in unambiguously recovered. Chains are still in the cross-over regime and the limiting scaling law, with exponent $2\cdot\nu-1$, is not yet valid (see figure 14). However, assuming that for $\omega$ sufficiently small the ratio $\langle \mathbf{R}_e^2 \rangle / \langle \mathbf{R}_e^2 \rangle^*$ is only a function of $L/l_T$ we have built a master curve by horizontally shifting curves on a log scale for various $\omega$ on the top of a reference curve ($\omega = 1.00$), trying only to superimpose values at larger $L$. The shift factor enable the calculation of $l_T$ for each $\omega$ assuming $l_T$ at $\omega = 1.00$ being arbitrary defined as the intersection of the limiting power law with the horizontal line $\langle \mathbf{R}_e^2 \rangle / \langle \mathbf{R}_e^2 \rangle^* = 1$ ($l_T/d \approx 20$, see figure 14(a)). Results are presented in figure 15 where data of figure 14 are plotted but as a function of $L/l_T$. Except for the highest $\omega$, data superimposes reasonably well and the cross-over can be described with the following phenomenological equation:

$$\frac{\langle \mathbf{R}_e^2 \rangle}{\langle \mathbf{R}_e^2 \rangle^*} = \left[ 1 + a \cdot (L/l_T)^{1/2} + (L/l_T) \right]^{2\cdot\nu-1} \quad (28)$$

with a = 0.3. This cross-over is large and takes place over about four decades (10$^{-2}$ << $L/l_T$ << 10$^2$). Figure 16 shows the evolution of $l_T/d$ as a function of $l_p/d$. The theoretical behavior predicted by Schaeffer et al. [77] is well recovered (see equation (27)).

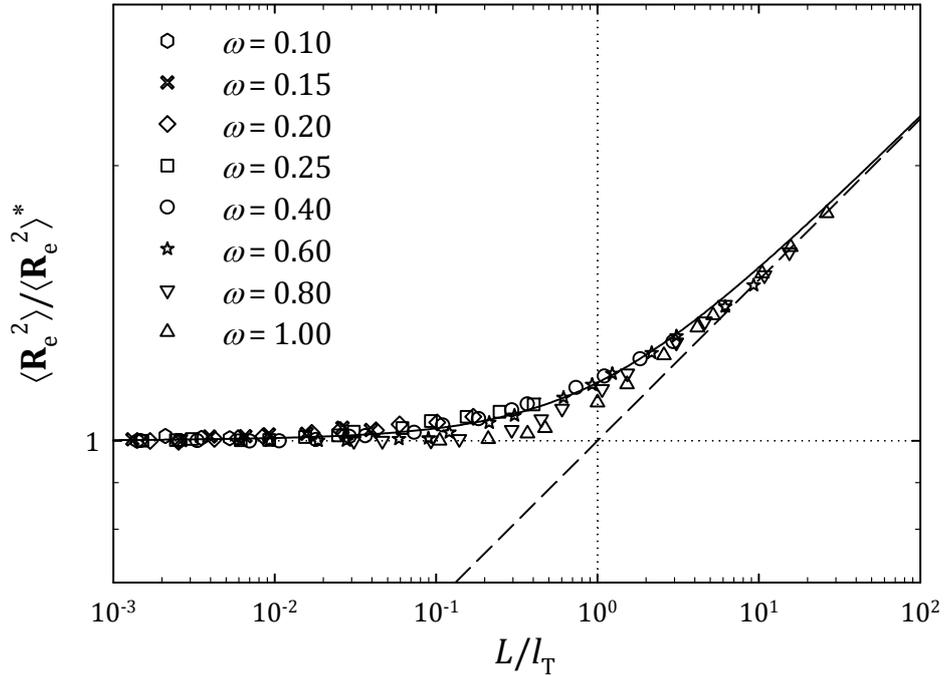

FIG. 15. Same data as in figure 15 but as a function of $L/l_T$ for various $\omega$ as indicated in the figure. The solid lines represent equation (28) with $a$ = 0.3. The dashed line has a slope 2·$\nu$-1 = 0.176. Dotted lines are construction lines.

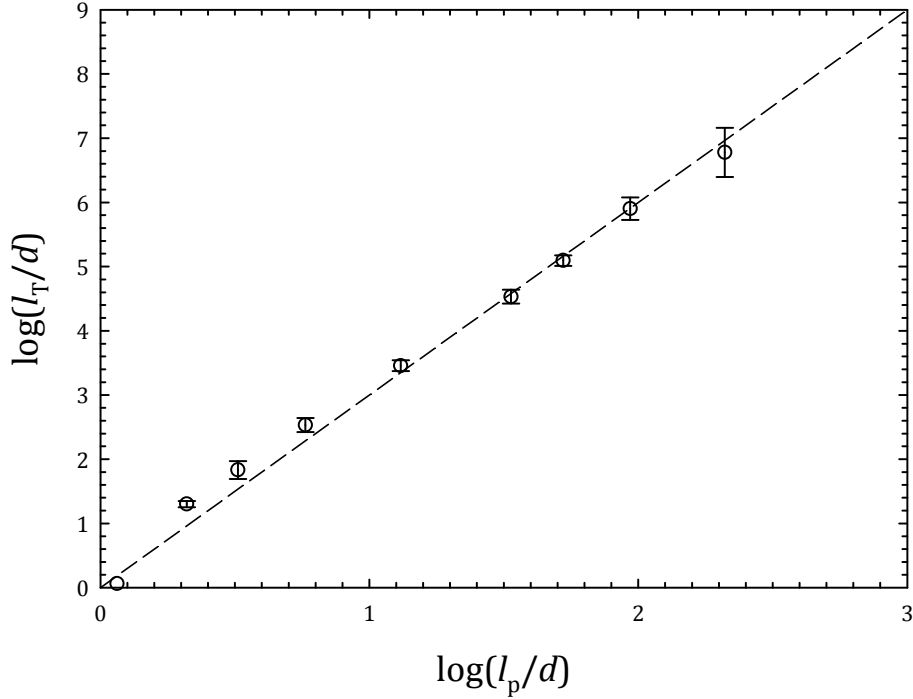

FIG. 16. Evolution of log($l_T/d$) as a function of log($l_p/d$) for patchy polymer chains with excluded volume interactions. The dashed line has a slope of 3 as predicted by Schaeffer et al [77].

Figure 17 compares the evolution of $\langle \mathbf{R}_e^2 \rangle/d^2$ as a function of $L/d$ with $\omega = 0.60$ for a self avoiding and an ideal chains. Three distinct behaviors are clearly visible when excluded volume effects are taken into account. For $L \ll l_p$ the chain is in the rigid domain; for $l_p \ll L \ll l_T$ it is flexible but with an ideal behavior and for $l_T \ll L$ it is flexible and swollen. Finally figure 18 shows the evolution of the ratio $\langle \mathbf{R}_e^2 \rangle/(6 \cdot \langle R_g^2 \rangle)$ as a function of $L/l_T$ for self-avoiding PPC. It is clear, from the figure, that the ratio for the most locally flexible chains tends toward the expected value given by renormalization group theory [66]: $\langle \mathbf{R}_e^2 \rangle/(6 \cdot \langle R_g^2 \rangle) = 1.0504$. While for the most locally rigid ones, largest chains generated still have small values of $L/l_T$ and are yet in the intermediate regime. One can reasonably think that they would reach the correct behavior if they were sufficiently long ($L/l_T \gg 1$).

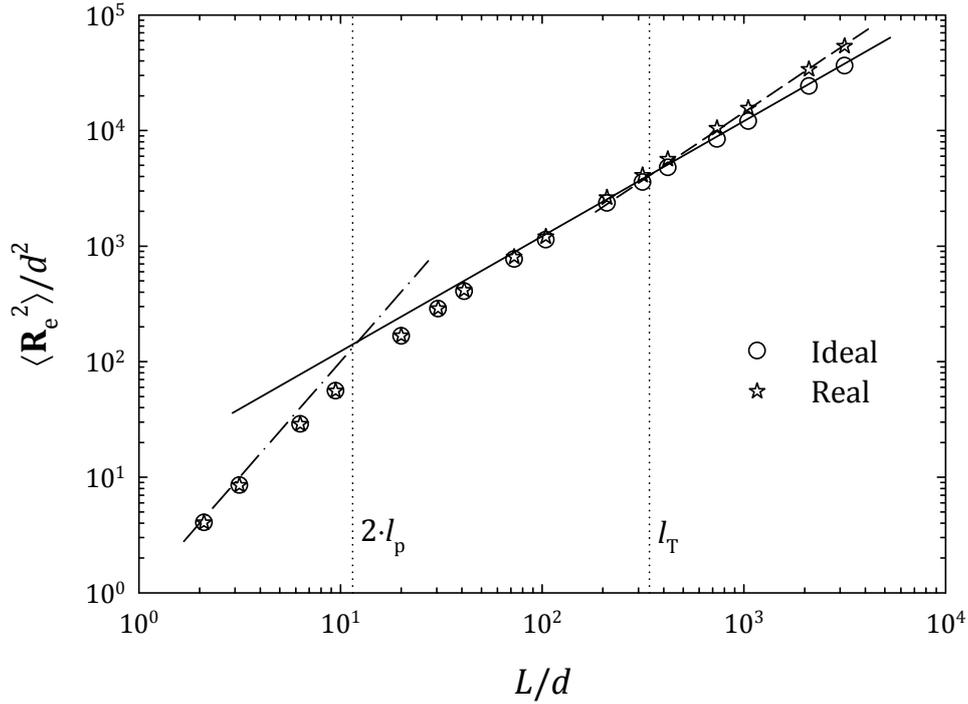

FIG. 17. Evolution of $\langle \mathbf{R}_e^2 \rangle / d^2$ as a function of $L/d$ on a log-log scale for a patchy polymer chains with $\omega = 0.60$. The dashed-dotted line has a slope of 2, the solid line a slope of 1 and the dashed line a slope of $2 \cdot \nu = 1.176$. Vertical dotted lines show respectively, on the left, twice the persistent length equivalent to the Kuhn length, and on the right, $l_T$. Error bars are smaller than the symbol size.

The PPC model appears to be in very good agreement with theoretical predictions for static properties of self avoiding polymer chains. It is an off-lattice model that enables a very convenient tuning of the local chain rigidity by playing on the patch angle. We did study other static properties like the local persistent length, $l_p(k)$, defined as the projection of the local bond vector $k$ on the end-to-end vector. This quantity has been introduced by Yamakawa [78] and theoretically investigated by Schäfer and Elsner [79]:

$$l_p(k) = \frac{\langle \mathbf{r}_k \cdot \mathbf{R}_e \rangle}{\langle l_b \rangle} \quad (29)$$

It has received much attention recently [75, 80, 81]. Our results concerning $l_p(k)$ are beyond the scope of that paper and will be published elsewhere.

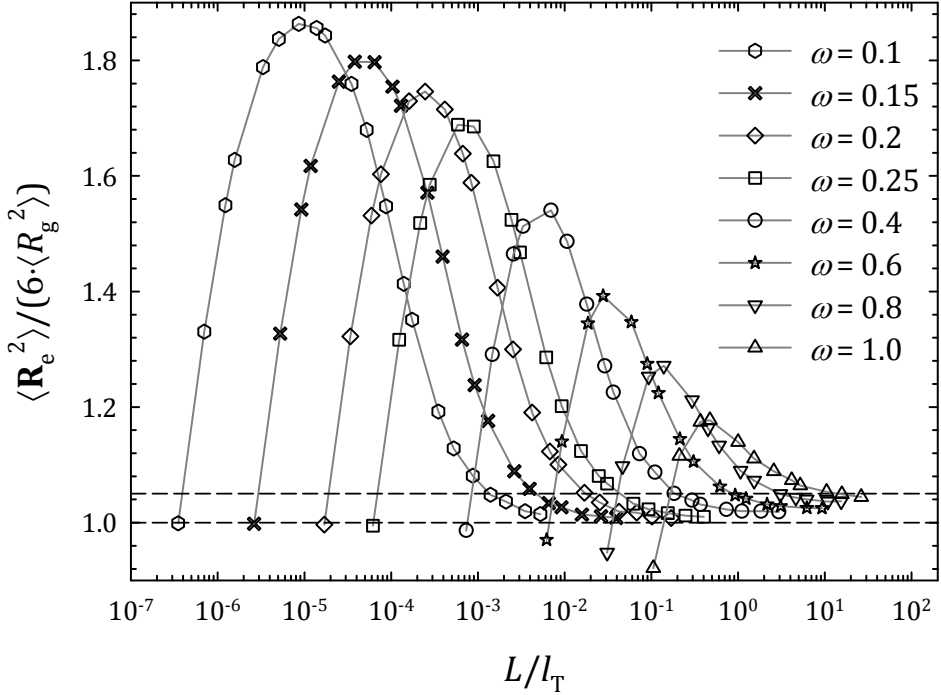

FIG. 18. Evolution of the ratio $\langle \mathbf{R}_e^2 \rangle / (6 \cdot \langle R_g^2 \rangle)$ as function of $L/l_T$ for real patchy polymer chains with various cone angle as indicated in the figure. Solid lines are guides to the eyes. The lower dashed line ($y = 1$) represents the expected values for long ideal chains while the upper one ($y = 1.0504$) represents the asymptotic behavior expected for long flexible self avoiding chains [66].

### E. Dynamical properties of self-avoiding patchy polymer chains using PBCD

#### 1. Translational diffusion

Figure 19 shows the evolution of the MSD of the center of mass and of an average monomer for a self-avoiding polymer chain with $m = 20$ and $\omega = 1.0$ at a given $s_T/d = 0.0131$. As expected at short times, monomers diffuse freely before they start to feel the constraint of their bounded neighbors. This is one of the advantages of our model to be able to reproduce correct dynamics at short times compared to conventional MD methods. Then monomers exhibit an intermediate behavior before merging with the one of the center of mass above a characteristic time proportional to $\tau_{max}$. As we are performing Rouse dynamics we expect, for long chains, a slope close to 0.5 in that regime. This seems reasonably well recovered in figure 19 considering that we have used a small chain with $m = 20$. It can be noted that the merging occurs when the MSD of the center of mass is of the order of the square end to end distance of the chain. Monitoring the MSD of the center of mass gives a direct access to $D_m^T$ (see equation (13)). This can be done very accurately even after short running times. In figure 20 we plot $D_m^T$ as a function of $\omega$ for self avoiding chains with various $m$. As expected in the framework where hydrodynamic interactions are ignored the local flexibility, while changing the chain size, plays no role on translational dynamics of the center of mass. $D_m^T$ is insensitive to $\omega$ and $D_m^T/D_1^T = 1/m$ (see figure 21). The total friction felt by the chain is simply the sum of individual frictions felt by monomers.

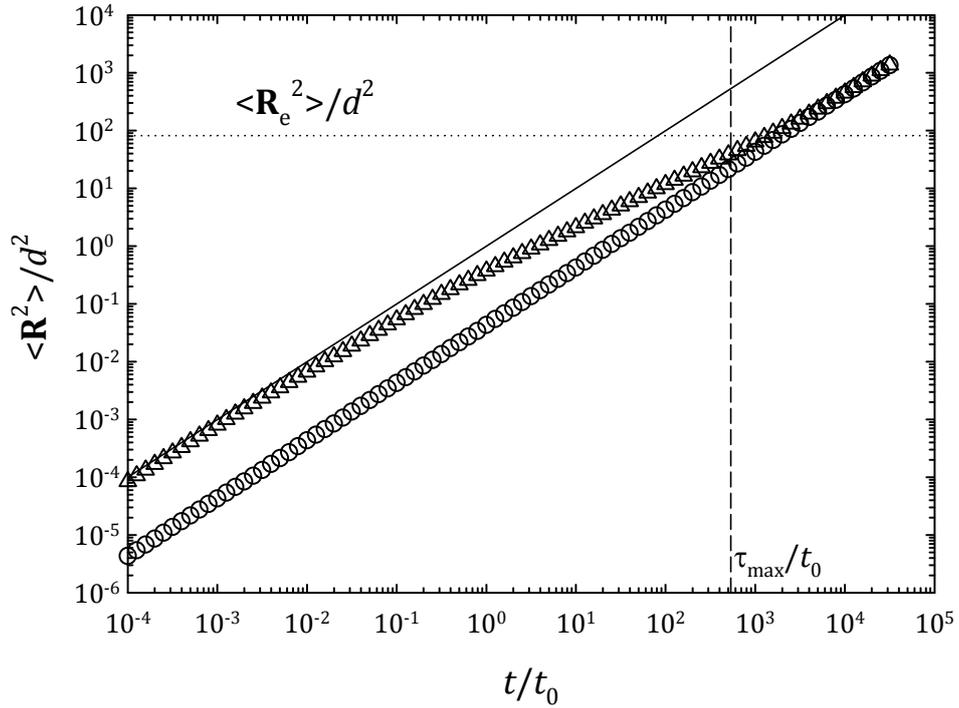

FIG. 19. Evolution of a polymer chain MSD, for an average monomer (triangles), and for the center of mass of the chain (circles). The measurement is done on self-avoiding polymer chain with $m = 20$ and $\omega = 1.0$ and $s_T = 0.0131$. The solid line ($y = x$) represents the expected behavior for a free monomer. The vertical dashed line represents $\tau_{max}/t_0 \approx 535$ measured from the end to end correlation function (see figure 3). The horizontal dotted line indicates the value of the mean square end to end distance of the chain ($\langle \mathbf{R}_e^2 \rangle / d^2 \approx 81.9$).

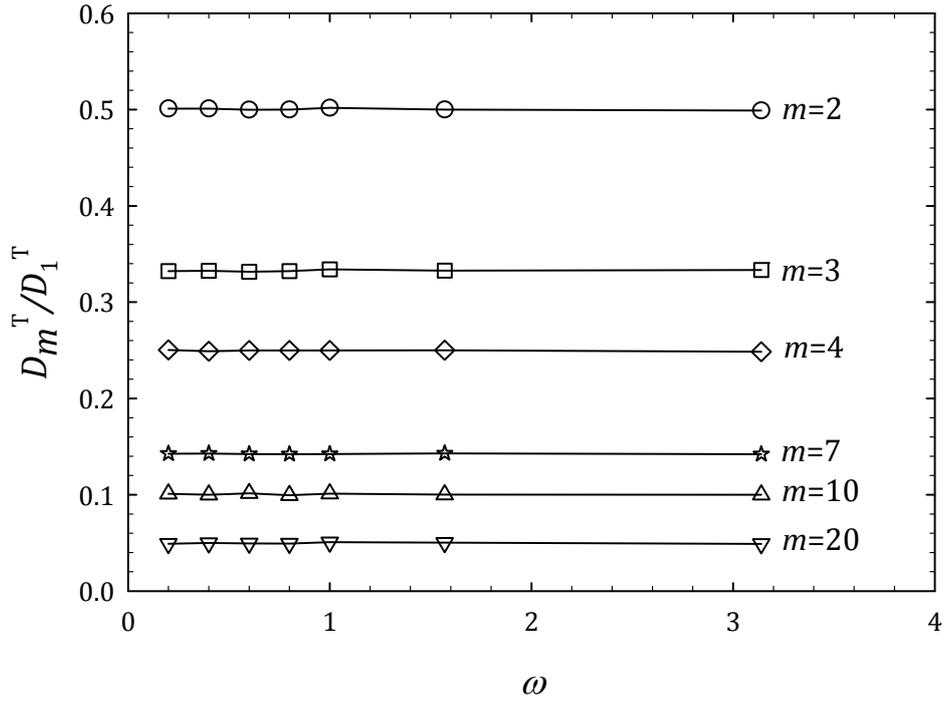

FIG. 20. Evolution of the ratio $D_m^T/D_1^T$ as a function of $\omega$ for various $m$ as indicated in the figure. Solid lines are guides to the eyes.

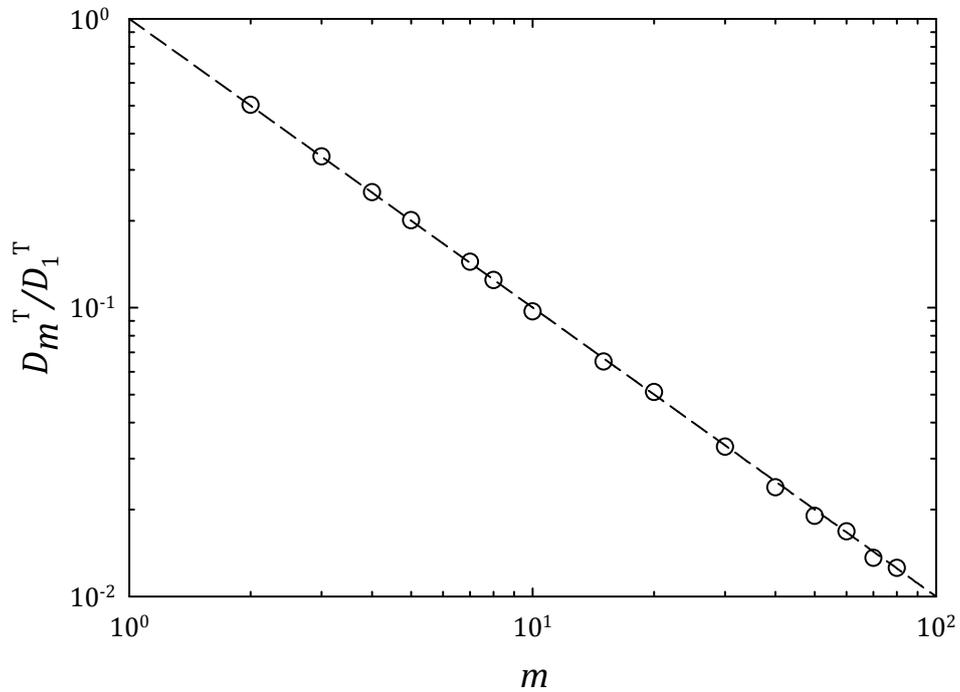

FIG. 21. Evolution of $D_m^T/D_1^T$ as a function of $m$. The dashed line represents the equation $y = 1/x$.

## 2. Rotational diffusion

Unlike translational dynamics of the center of mass, the rotational diffusion is highly time consuming (see Sec. II.B.) and is sensitive to the local flexibility of the chain. This is observed in figure 22 in comparison with figure 20 where we plot $\tau_{max}/t_0$ as a function of $\omega$ for various chain lengths. For a given $m$, $\tau_{max}$ increases as the chain becomes more rigid and two different behaviors are expected in the limit of rigid or flexible chains:

$$\begin{cases} \tau_{max} \propto m^3 & X \ll 1 \\ \tau_{max} \propto m^{2\cdot\nu+1} & X \gg 1 \end{cases} \quad (30)$$

with, again, $X = L/l_p$. To verify these predictions, we have plotted in figure 23 $\tau_{max}/(t_0 \cdot m^3)$ as a function of $X$ for all simulated chains. Data reasonably superimposes on a master curve considering longest chains only have 20 beads. There exist systematic deviations due to the finite extent of the chains, but results are in good agreements with theoretical predictions given by equation (30). The longest and most flexible self avoiding chains seem to behave with the correct scaling. Using the longest chain we can study ($m = 20$ see Sec. II.B.) we compare the shape of $C(t)$ as a function of $t/\tau_{max}$ for two different $\omega$. It is clear from figure 24 that as the chain becomes locally more flexible $C(t)$ behaves similarly to equation (14) while as it is more rigid it loses relaxation modes and $C(t)$ comes closer to a single exponential relaxation.

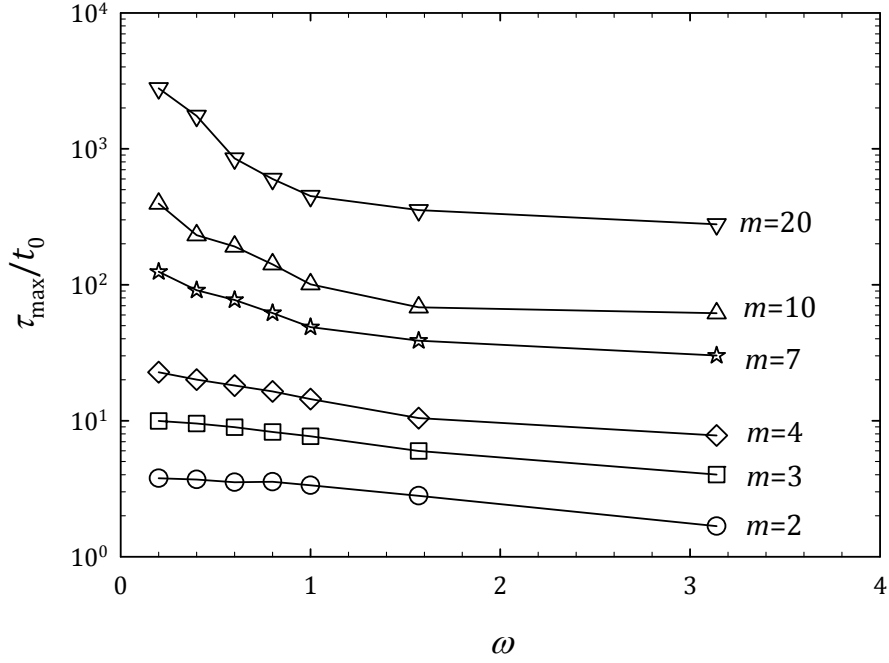

FIG. 22. Evolution of $\tau_{max}/t_0$ as a function of $\omega$ for self avoiding chains with various $m$ as indicated in the figure. Error bars are of the order of symbol size. Solid lines are guides to the eyes.

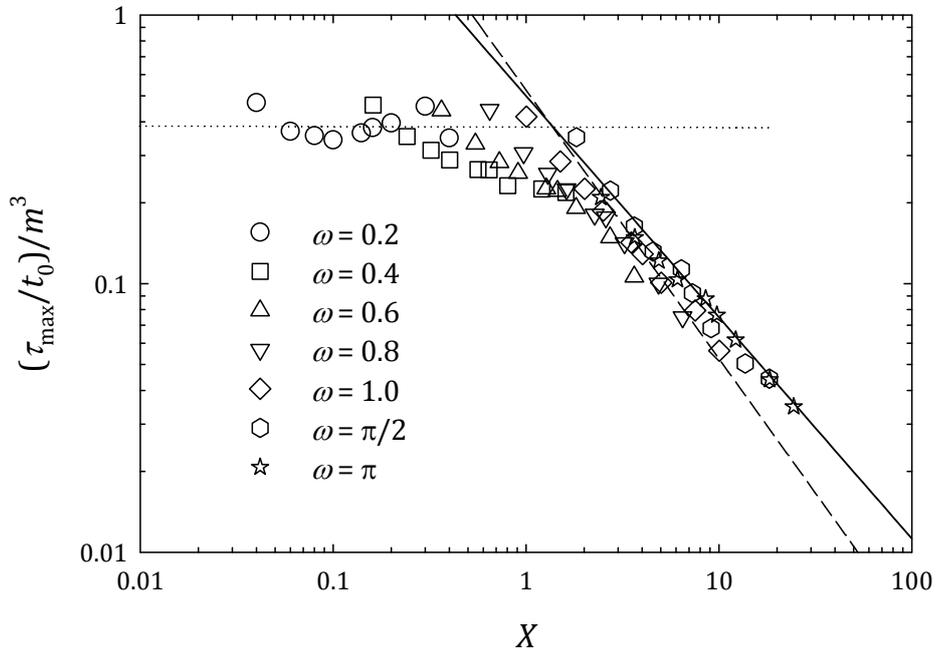

FIG. 23. Evolution of $\tau_{max}/(m^3 \cdot t_0)$ as a function of $X = L/l_p$ for self avoiding chains with various $\omega$ as indicated in the figure. Error bars are of the order of symbol size. The horizontal dotted line figures the plateau expected for rod like chains ($X \ll 1$). The dashed line has a slope of -1 and the solid line a slope of $2 \cdot \nu$-2 = -0.824. They represent the expected behavior for long flexible ideal or swollen chains ($X \gg 1$).

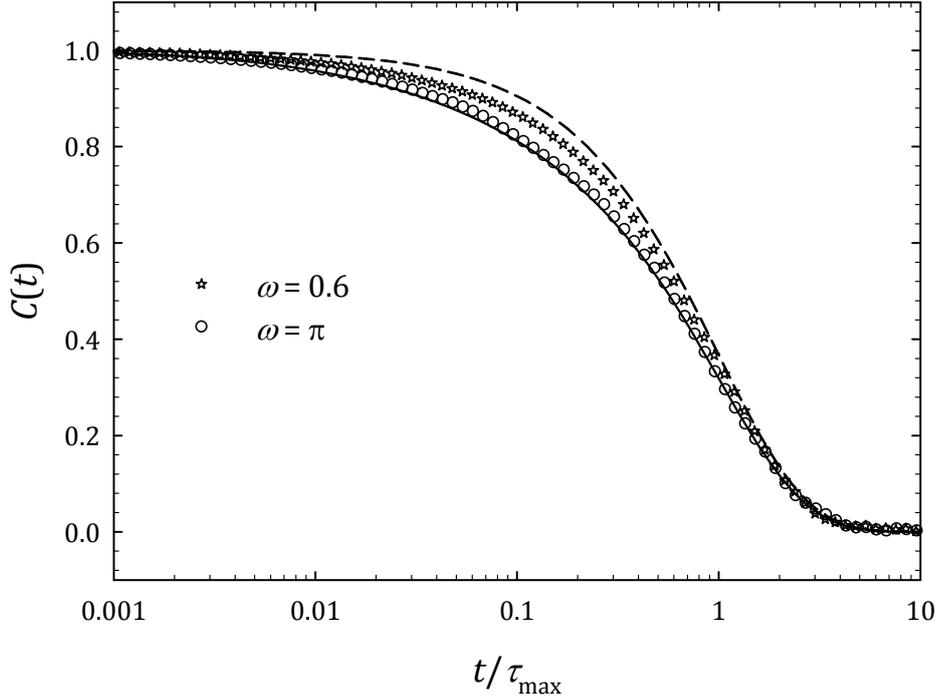

FIG. 24. Evolution of $C(t)$ as a function of $t/\tau_{max}$ for a self avoiding chains with $m = 20$, $s_T = 0.0131$ and two different $\omega$ as indicated in the figure. Error bars are of the order of symbol size. The dashed line shows the single exponential behavior expected for a rigid rod while the solid line figures the expected behavior for a long flexible self avoiding chain (see equation (14)).

## IV. PRELIMINARY RESULTS ON STEP-GROWTH POLYMERIZATION. FORMATION OF OUT OF EQUILIBRIUM ARRESTED STRANDED GELS

This part is dedicated to some preliminary results about the competition between polymerization and phase separation. It demonstrates the efficiency of our algorithm to investigate arrested out of equilibrium structures.

Here, we introduce a secondary isotropic SW potential coupled to the patches. Equilibrium properties of such a system have been investigated by Liu et al. [82] using MC simulations and a thermodynamics perturbation theory approach. They limit their study to one possible bond per patch using geometrical constraints to be consistent with the theoretical assumptions [83]. They focused on the influence of the number of patches and their coverage on the resulting phase diagram and showed that the second virial coefficient is a scaling parameter for a generalized law of corresponding states [36, 84]. In this study we concentrate on the competition between an irreversible patchy polymerization (highly directional) coupled with an isotropic reversible interaction that mimics the quality of the solvent. For simplicity, the range of the isotropic SW is the same as that for patches ($\varepsilon = 0.1$). In a bad solvent, patch free particles will have a tendency to self associate and even phase separate. The resulting SW fluid has already been extensively studied using BCD [24-26]. Its adhesiveness can be characterized by the attractive part ($B_{att}$) of its reduced second virial coefficient ($B_2$). We speak about a reduced quantity that has been divided by the particle volume to obtain dimensionless numbers. For a SW fluid we have:

$$B_2 = 4 - B_{att} \tag{31}$$

with

$$B_{att} = 4 \cdot \left[ \exp\left(-\frac{u_0}{k_B \cdot T}\right) - 1 \right] \cdot \left[ (1+\varepsilon)^3 - 1 \right] \tag{32}$$

$B_{att}$ combines in a single parameter the range and the energy of the interaction. It has been shown that in the limit of small ranges ($\varepsilon \ll 1$), the phase diagram is no more influenced by the precise shape of the potential and a unique phase diagram is obtained in the plane ($B_{att}$, $\phi$) [84]. Previously we have shown that systems with $\varepsilon = 0.1$ can be considered in that limit [25, 85]. When $B_{att} = 0$, the system is in good solvent conditions but as $B_{att}$ increases, its thermodynamic quality decreases and favors isotropic reversible aggregation in the solution. When $B_{att} = 4$, the hard core repulsion is balanced by the attractive part of the potential and $B_2 = 0$. This corresponds to the Boyle temperature of the fluid. To some point, it can be considered to be close to theta conditions for the polymer chain [86]. Figure 25 (adapted from [26]) shows the phase diagram of a SW fluid. To trigger a spontaneous crystallization it is necessary to work below the metastable liquid-liquid binodal. Phase separation above that line is nearly impossible without introducing crystalline seeds in the solution. Snapshots in figure 25 compare a system with $\phi = 10\%$ at the same time, above ($B_{att} = 4$) and below ($B_{att} = 12$) the liquid-crystal binodal. In the first case the system is homogeneous while in the second one, we observe crystallization that takes place in dense liquid droplets.

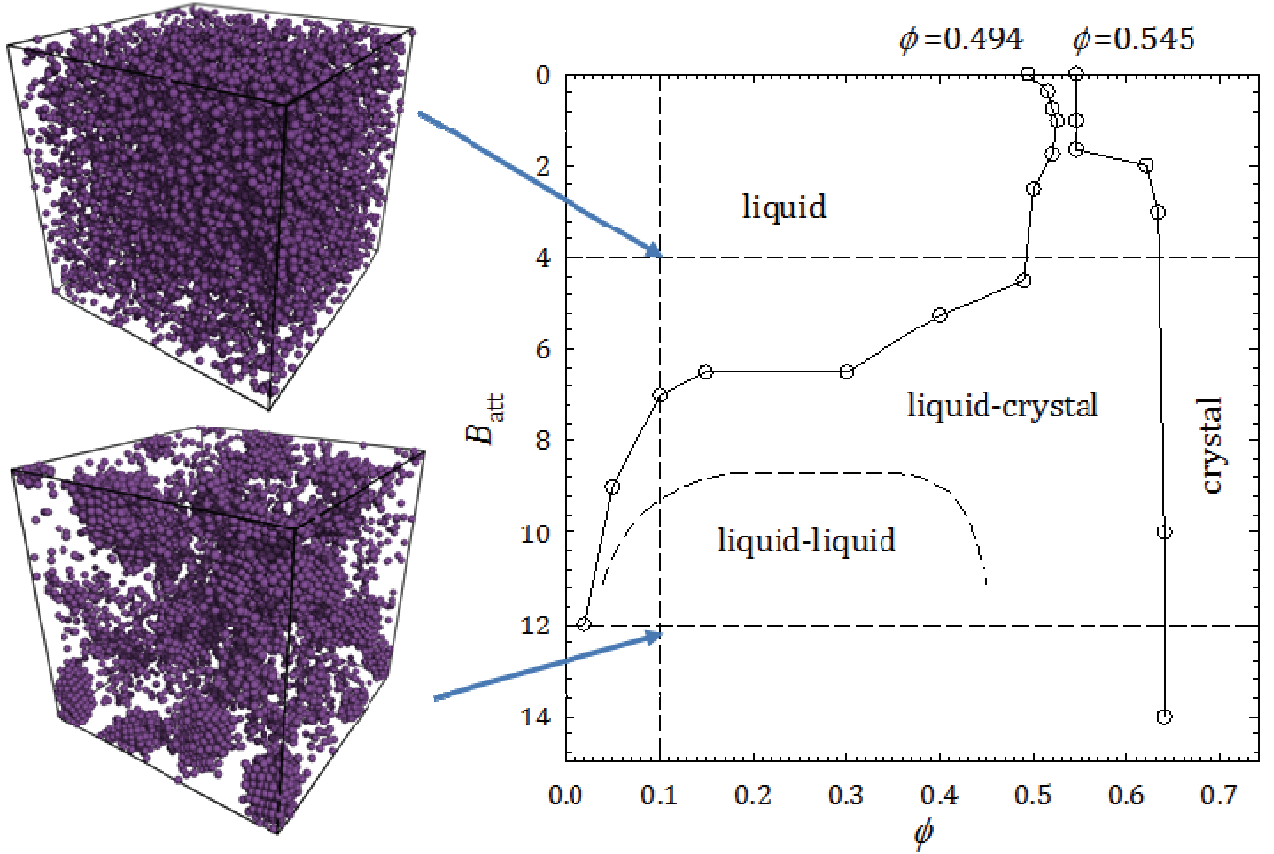

FIG. 25. Adapted from [26]. Phase diagram of a SW fluid with relative range $\varepsilon = 0.1$. The attraction strength is expressed in terms of the attractive part of the reduced second virial coefficient, see text. The circles indicate the binodal of the crystal-liquid phase separation. The dash line curve shows the metastable liquid-liquid binodal. Below that curve, crystallization takes place in dense liquid droplets. At $B_{att} = 0$ we have a simple hard sphere fluid with a liquid-crystal mixture between $\phi = 0.494$ and $\phi = 0.545$. The vertical and both horizontal dash lines indicate respectively $\phi = 0.1$ and $B_{att} = 4$ or $12$ where the influence of the isotropic SW potential has been tested on the irreversible patchy aggregation. Snapshots of the system for both conditions are represented on the left side of the figure. Pictures have been taken at $t/t_0 = 1073$ starting in both cases from a random distribution of particles. The simulation box contains 9674 particles and has a linear size $L_{box}/d = 37$.

We introduce now patchy irreversible aggregation to take into account the step-growth polymerization process in both solvent conditions. An infinite patchy potential is simply added to the finite isotropic one. For some convenience we have used patches with angle $\omega = 0.8179$, so that polymerization kinetics is not too slow and takes place within similar time scales as the phase separation. In that case, patches cover a total amount of 32% of the particle surface. In good solvent ($B_{att} = 0$, self avoiding chains) or close to the theta condition ($B_{att} = 4$, ideal chains), isolated chains built with that cone angle would be rather flexible with a persistent length around 3 bonds (see figure 14). Figure 26 shows two snapshots of the polymerizing system at the same time and for both solvent conditions. Individual chains are shown with different colors to figure out their length. Figure 26(a) ($B_{att}$ = 4) shows a semi-dilute solution of entangled polymer chains with a number average degree of polymerization ($m_n$) around 219. Initially, theta conditions enhance slightly polymerization kinetics by favoring contacts between monomers compared to a good solvent ($B_{att}$ = 0) but at larger times both kinetics are nearly identical and polymerizations in good or theta solvents look very similar. At $B_{att}$ = 12 (figure 26(b)), things appear very different. Chains are in average much smaller ($m_n$ = 102) and the system exhibits a remarkable structure. It is a network made of strands and nodes. Strands gather several "stretched" chains which dip into entangled nodes where their conformation is rather "globular".

(a)          (b)

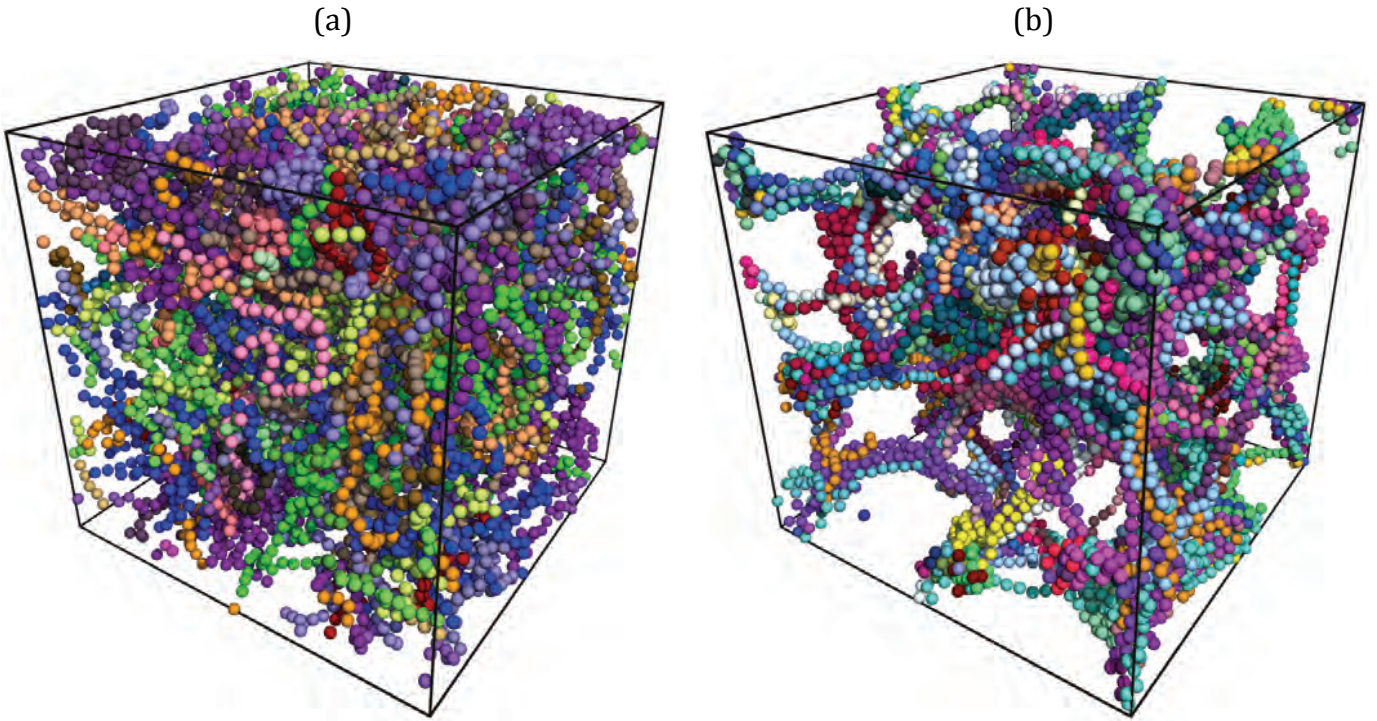

FIG. 26. Snapshots of the previous system undergoing a concomitant step-growth polymerization with $\phi$ = 0.1, $\varepsilon$ = 0.1 and $\omega$ = 0.8179 at (a) $B_{att}$ = 4 and (b) $B_{att}$ = 12. As in figure 25, pictures have been taken at $t/t_0$ = 1073. In both cases, the simulation starts from a random distribution of particles. The simulation box contains 9674 particles and has a linear size $L_{box}/d$ = 37. Individual chains are represented with different colors.

Figure 27 shows the system at a longer polymerization time ($t/t_0 = 5378$). Plotting only bonds between monomers (figure 27(b)), reveals the structure more clearly. Chains are completely out of equilibrium and should adopt globular conformations in such a bad solvent ($|T_\theta - T|/T_\theta \approx 0.4$, where $T_\theta$ is the theta temperature). Moreover, no major differences exist between figure 26(b) and 27(a) while the reaction time has been multiplied by five. The system looks arrested, trapped in a metastable configuration.

(a)  (b)

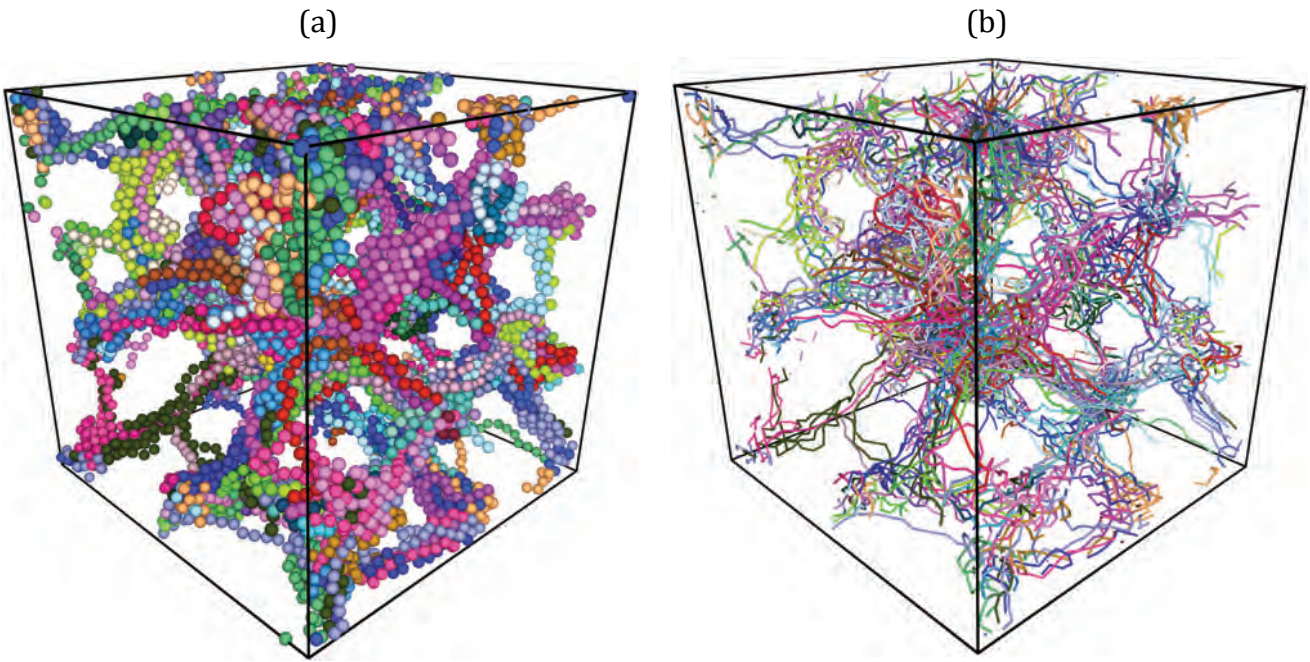

FIG. 27. Snapshots of the previous system undergoing a step-growth polymerization with $\phi = 0.1$, $\varepsilon = 0.1$, $\omega = 0.8179$ and $B_{att} = 12$. Pictures have been taken at $t/t_0 = 5378$. Individual chains are represented with different colors. In (b) monomers are not shown and only colored bonds between them are displayed with the same color code.

To verify whether it is arrested or not, we have first monitored the degree of reaction of the polymerization, $x$, which is simply the fraction of reacted patches. Plotting $1-x$ or $m_n$ as a function of the time on a log-log scale reveals clearly the difference of kinetics between $B_{att} = 12$ or 4 (see figure 28). The reaction appears to have stopped for $B_{att} = 12$ while it continues to slowly evolve at $B_{att} = 4$. The system in figure 27 is "chemically" arrested. We have also monitored dynamics in both cases by plotting the MSD of an average monomer during the polymerization reaction for two different waiting times ($t_w$). Figure 29 shows the MSD of an average monomer for both solvent conditions with $t_w/t_0 = 0$ and $t_w/t_0 = 5378$. For $t_w/t_0 = 0$ monomers exhibit free diffusing dynamics at short times which becomes sub-diffusive as the reaction proceeds. At large times, for $B_{att} = 12$ a plateau is reached while for $B_{att} = 4$ dynamics is more compatible with the one of a Rouse polymer chain at a time scale smaller than the Rouse time (see for example [72]). Restarting to monitor the MSD at $t_w/t_0 = 5378$ we also notice a clear difference of behavior. At $B_{att} = 12$ a monomer endures in average a displacement of its diameter over a period of time equal to around $5000 \cdot t_0$. At $B_{att} = 4$ monomer dynamics is much faster and seems to converge toward the one of a monomer in a Rouse chain at a time scale smaller than the Rouse time. For $B_{att} = 12$, the reduction of dynamics at larger time scales supports the argument of a dynamically arrested system. Only "breathing motions" of the stranded gel remain visible (see both movies in supplementary materials [87], they illustrate both kind of dynamics) and the system can reasonably be considered chemically and dynamically arrested.

We have checked the effect of the size of the simulation box using a bigger box and found that the resulting structure does not depend on the system size. Figure 30 shows two snapshots taken at the same time ($t/t_0 = 603$) for $L_{box}/d = 100$ (containing 190985 monomers) and $L_{box}/d = 37$ (containing 9674 monomers) which corresponds to $\phi = 0.1$ and at $B_{att} = 12$. We do not observe appreciable differences in the system and both systems reach the same polymerization degree at the same time. However, the computation time needed for the bigger system is about thirty times bigger and does not allow us to go much further.

Figure 31 shows the effect of concentration on the obtained network structure. When the concentration decreases, the correlation length of the system seems to increase and the strands become more stretched and contain a lesser number of chains. Whether there exists a critical concentration below which the network structure is not formed is still an open question. At a lower concentration, the polymerization is slower and the nodes may act as nucleation sites for micro phase separation.

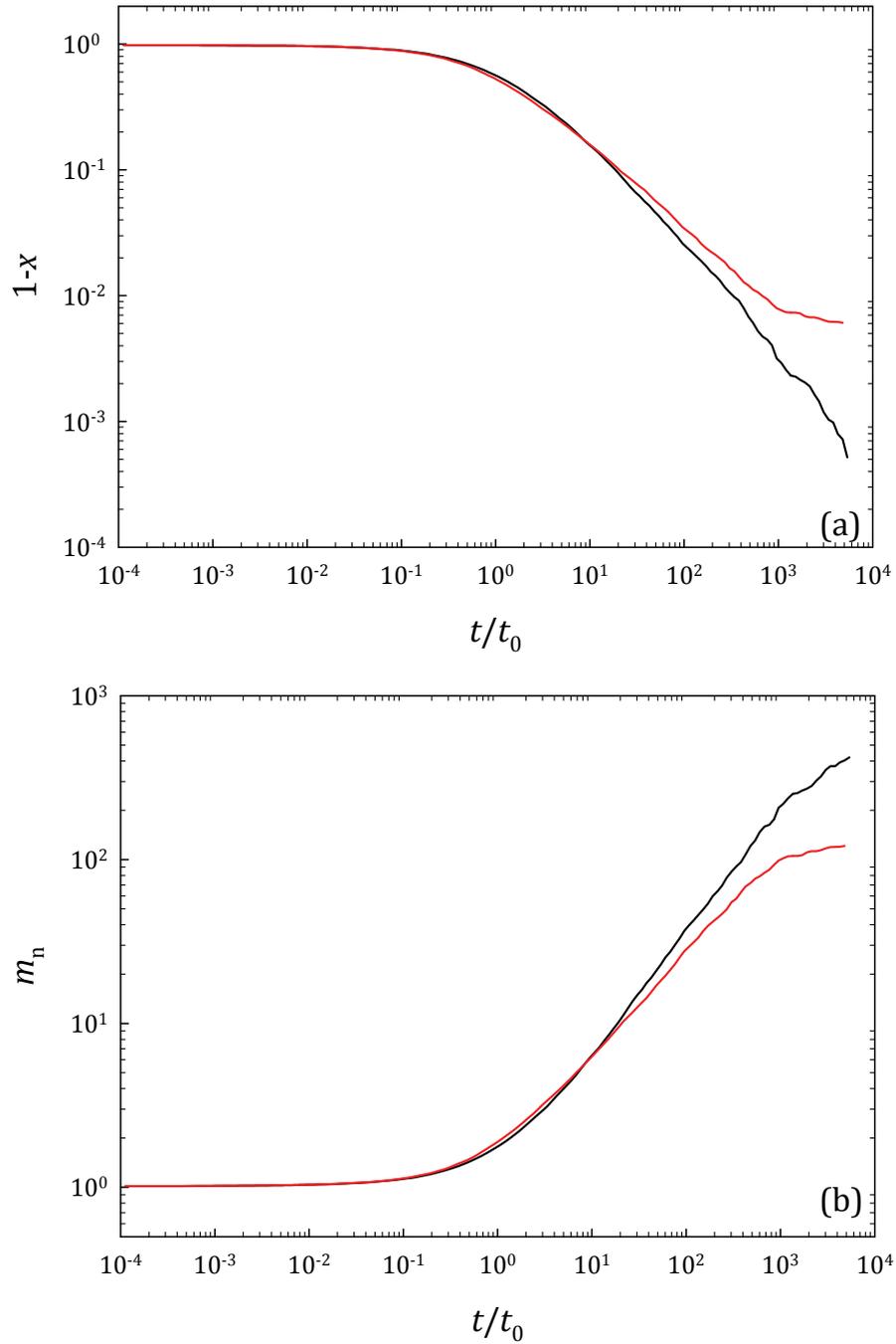

FIG. 28. Kinetics of polymerization for $\phi = 0.1$ and $\omega = 0.8179$. (a) $1-x$ as a function of $t/t_0$ for $B_{att} = 4$ (black curve) and $B_{att} = 12$ (red curve). (b) $m_n$ as a function of $t/t_0$ for $B_{att} = 4$ (black curve) and $B_{att} = 12$ (red curve).

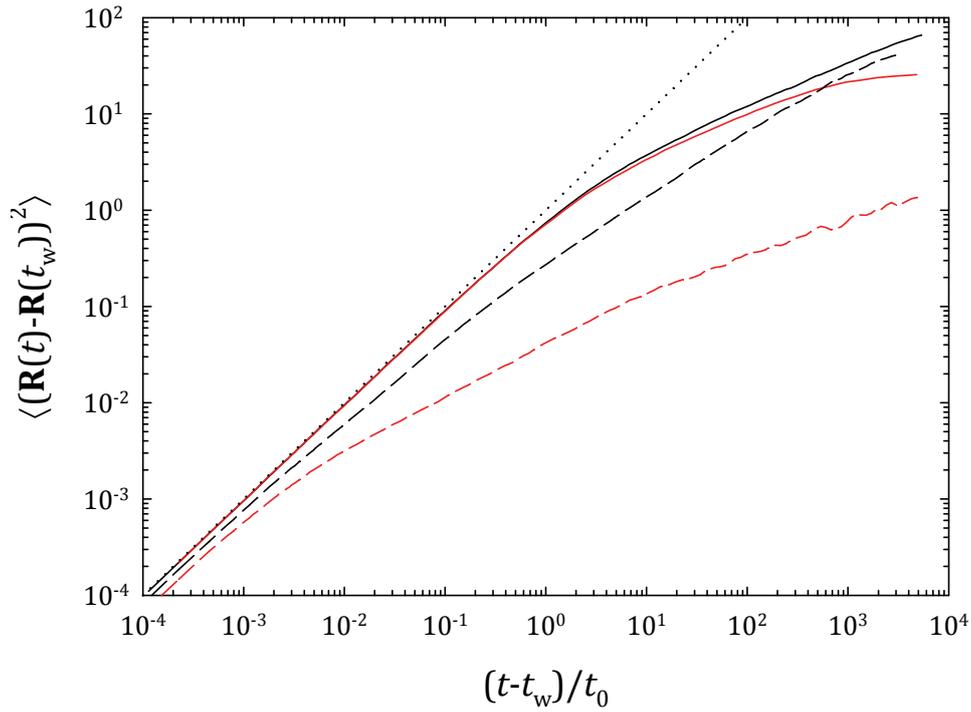

FIG. 29. Evolution of the MSD of an average monomer during the polymerization reaction for both solvent condition: $B_{att}$ = 4 (black curve) and $B_{att}$ = 12 (red curve). Full line curves have been obtained with $t_w/t_0$ = 0 and dashed lines with $t_w/t_0$ = 5378. For comparison, the dotted line shows the theoretical behavior of a free diffusing monomer.

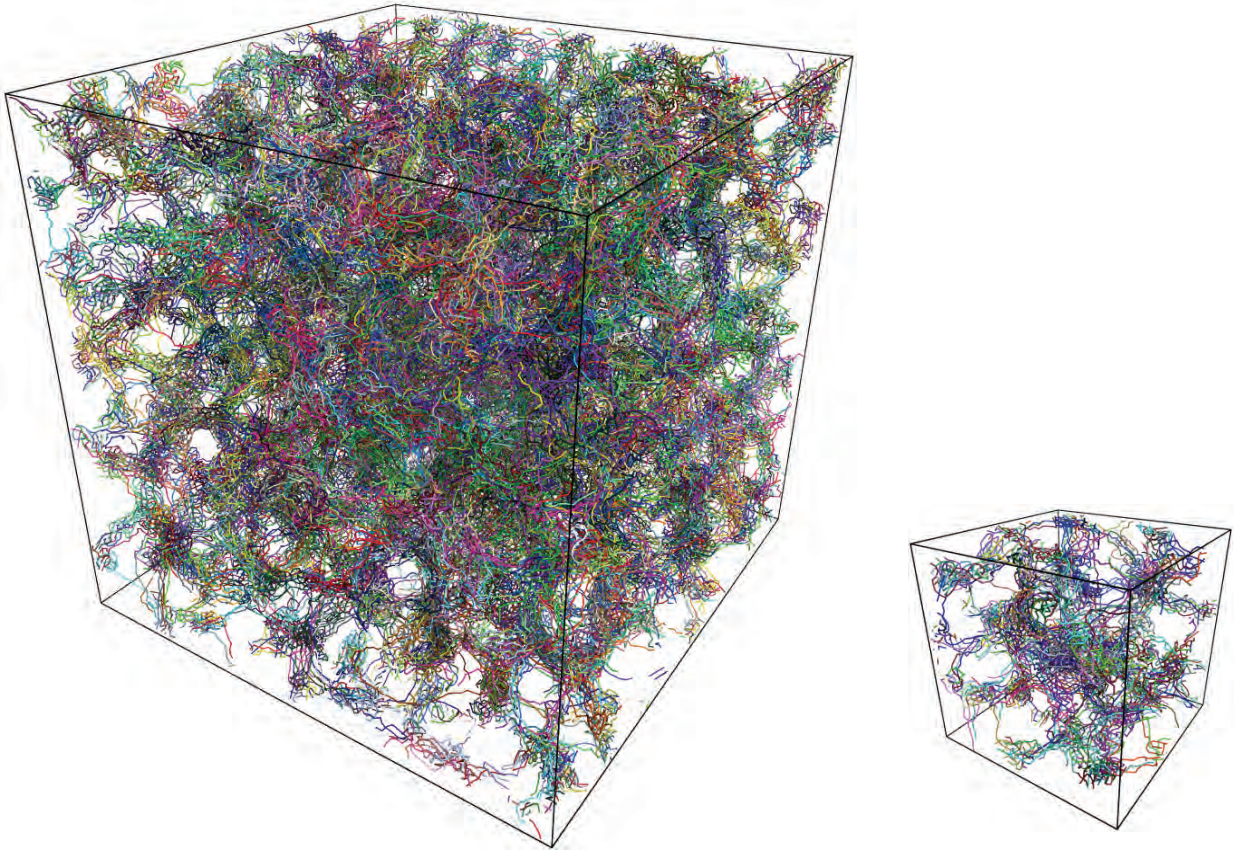

FIG. 30. Snapshots of a polymerizing system taken at $t/t_0 = 1073$ with $\phi = 0.1$ and $B_{\text{att}} = 12$ for (a) $L_{\text{box}}/d = 100$ (190985 monomers) and (b) $L_{\text{box}}/d = 37$ (9674 monomers). Length scale is the same for both figures. Bonds are represented with different colors for different chains.

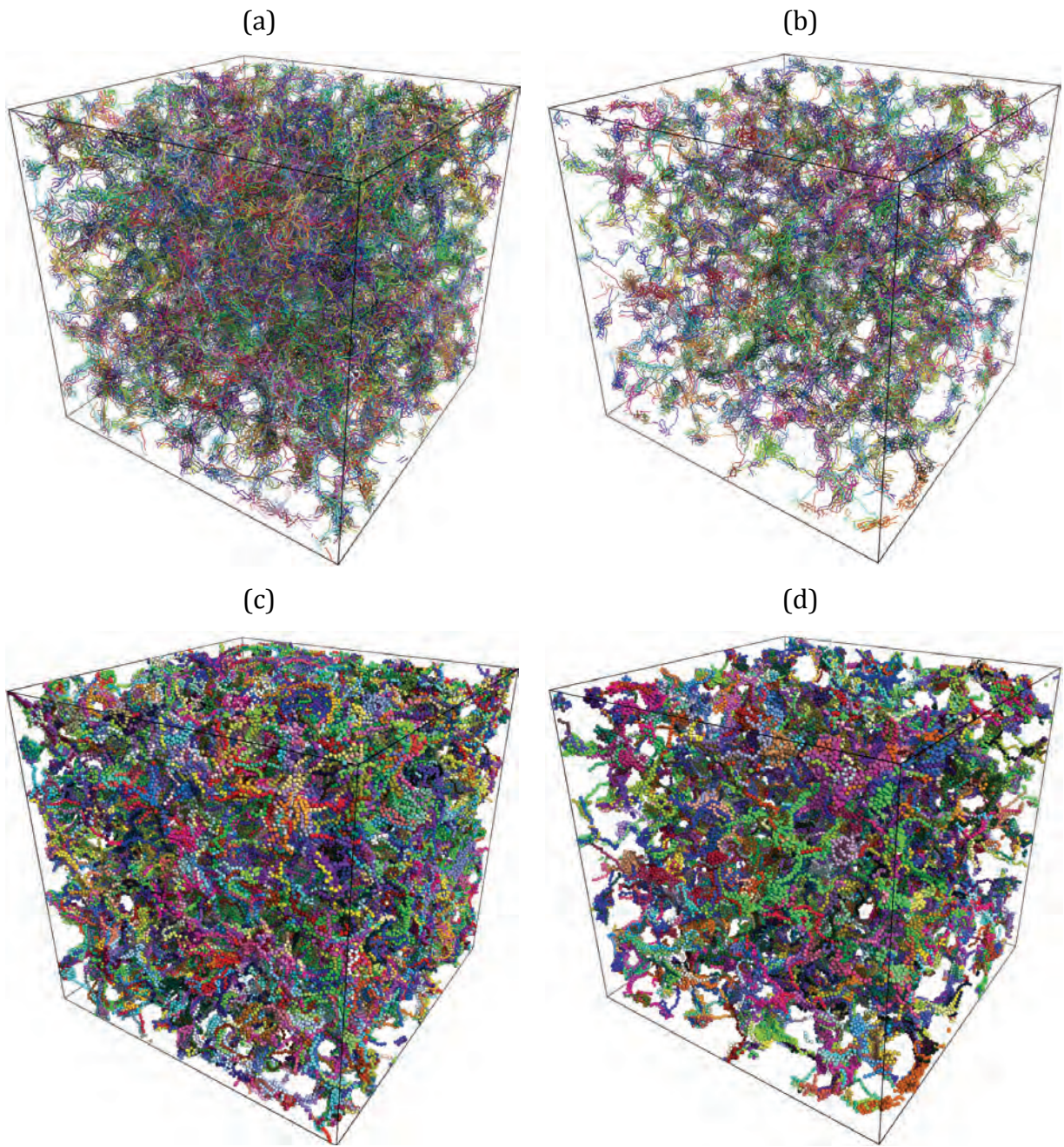

FIG. 31. Snapshots of a polymerizing system taken at $t/t_0 = 1073$ with $L_{box}/d = 100$ and $B_{att} = 12$ and two different concentrations. (a) $\phi = 0.10$, with bonds. (b) $\phi = 0.05$ with bonds. (c) $\phi = 0.10$ with monomers. (d) $\phi = 0.05$ with monomers. Bonds or monomers that belong to different chains have different colors.

## V. CONCLUSION

Finally to conclude we have seen that the PBCD algorithm applied to a single linear polymer chain in good solvent gives expected static and dynamical properties. Billions of small local random motions (translations and rotations), constrained by excluded volume interactions and the connectivity of the chain, lead to the correct collective dynamics in the framework where hydrodynamic interactions are ignored. It was also an opportunity, using the PPC model, to revisit the influence of the local flexibility on the bond correlation function and on the transition from ideal chain to swollen chain for self avoiding semi flexible polymers.

We would like to note here that PBCD is not specially designed to study dynamics of polymer chains but rather to focus on complex aggregation reactions involving both isotropic and directional potentials. Neither it is optimized to determine accurately equilibrium states if they exist and a Metropolis approach has to be preferred for that purpose. However PBCD is particularly suited to monitor the temporal evolution of particulate systems far from their equilibrium (like phase separating system with strong interactions) or out of equilibrium when they are kinetically driven by chemical reactions involving the formation of irreversible bonds. The example given in Sec. IV illustrates the ability of our algorithm to give relevant insights where classical tools (molecular dynamics or theoretical approaches) are inoperative. It enables us to study the competition between polymerization and phase separation that leads to some very interesting out of equilibrium structures like stranded gels. These structures can be viewed as arrested microphase separation resulting from a balanced competition between both effects. The solidity of the resulting 3d network results from a combination of irreversible bonds (by polymerization) and highly cooperative reversible interactions between polymer chains that form gel strands and nodes.

With this model we can study static, kinetic and dynamic properties of irreversible as well as reversible aggregation processes with varying the patch size, the strength of the anisotropic potential adding or not an isotropic potential for competition effects. This could provide more insights into the connection between gels and gas-liquid behavior of colloidal systems. We would like to extend our model to study physical systems like Janus particles but also the formation of calcium-silicate-hydrate (CSH) gels formed during cement hydration [88]. In the later case, it could be interesting to mix an aggregation process with the nucleation and growth of the basic building blocks of CSH gels [89].

**ACKNOWLEDGMENTS**

This research was performed thanks to a CNRS-INRA doctoral grant. We would like to thank Yvan Labaye (IMMM, Le Mans, France) for providing us with the spherical grid algorithm and figure 32(a). The computational resources of CESGA (Galicia, Spain) are gratefully acknowledged.

**APPENDIX A: the spherical grid**

A spherical grid is obtained starting from a regular icosahedron inscribed in the sphere with radius $v$. It is made of 12 vertices, 20 faces and 30 edges of length $a = v/\sin(2\pi/5)$. Giving $q$, a none zero positive integer, the grid is constructed by: transforming each of the original equilateral triangular faces into $q^2$ new equilateral triangular faces with edge-length $a/q$ and projecting each new vertex obtained onto the sphere. As there are $q$-1 new vertices per edge and $(q$-2$)\cdot(q$-1$)/2$ per face, the total number of directions (vertices) obtained is: $N_{\text{dir}} = 10\cdot q^2+2$. Figure 32(a) shows such a grid for $q=4$. For $q$ sufficiently large, the density of vertices on the surface is isotropic and $s_R$ is calculated as the square root of the average square distance between nearest-neighbors. Except the 12 original vertices that have 5, all have 6 neighbors. The rotational diffusion of the sphere is modeled by a random walk of **v** on the spherical grid with an average step size $s_R$. Figure 32(b) shows the evolution of $s_R/v$ as a function of $q$. As the distance between nearest-neighbors at the surface of the sphere is inversely proportional to the square root of the number of vertices, we have $s_R \propto 1/q$.

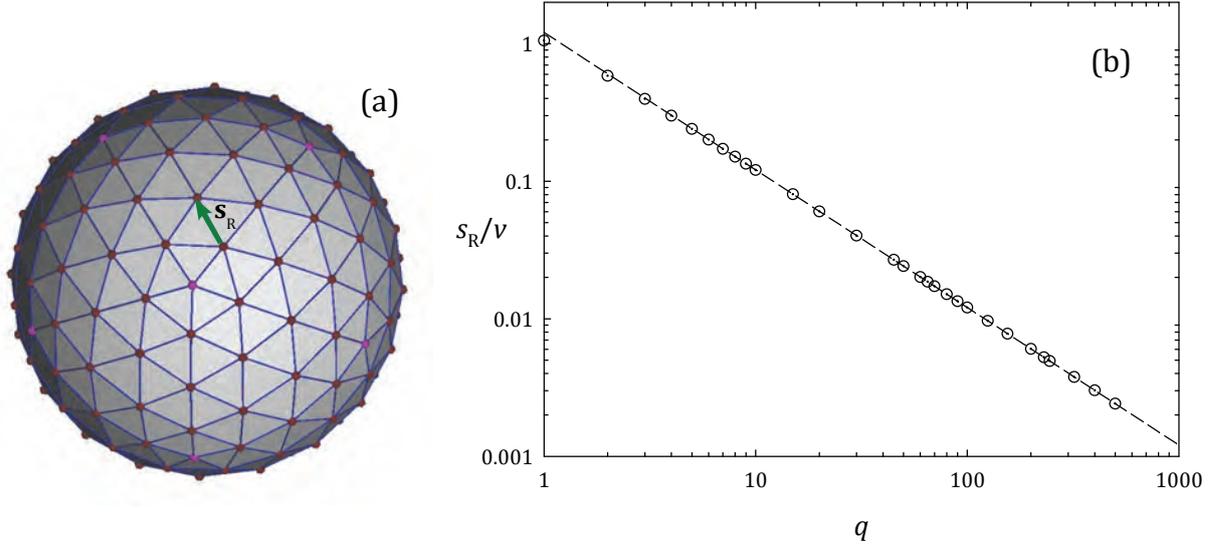

FIG. 32. (a) Representation of the spherical grid for $q = 4$ leading to $N_{dir} = 162$. (b) Evolution of $s_R/v$ as a function of $q$ on a log-log plot. The dashed line has a slope -1. See appendix A.

## APPENDIX B: averages bond parameters for the ideal PPC

For ideal PPC all average local quantities can easily be obtained by considering an uniform distribution of bonds within conic shells. We obtain:

$$\langle l_b \rangle^* = \sqrt{\langle \mathbf{r}_i^2 \rangle} = d \cdot \sqrt{\frac{3}{5} \cdot \frac{(1+\varepsilon)^5 - 1}{(1+\varepsilon)^3 - 1}} \tag{33}$$

$$\langle \cos\gamma \rangle^* = \frac{1+\cos\omega}{2} \tag{34}$$

$$\langle \cos\delta \rangle^* = \frac{(1+\cos\omega)^2}{4} \tag{35}$$

Star symbols indicate ideal quantities.

## APPENDIX C: average sizes for ideal FJC and FRC (see [71] for example)

Chains of *m* monomers are constructed with constant bond length $l_b$. For FJC model bond angles are arbitrary while for FRC they are constant equal to $\delta$. The average square end to end distance, $\langle R_e^2 \rangle$, and the average square radius of gyration, $\langle R_g^2 \rangle$, are quantities of interest that can be exactly calculated for both models. They are defined by:

$$\langle \mathbf{R}_e^2 \rangle = \sum_{i=1}^{m-1}\sum_{j=1}^{m-1} \langle \mathbf{r}_i \cdot \mathbf{r}_j \rangle = (m-1) \cdot l_b^2 + 2 \cdot l_b^2 \cdot \sum_{i<j}^{m-1} \langle \cos\theta(n) \rangle \tag{36}$$

where $\theta(n)$ is the angle between bond $\mathbf{r}_j$ and $\mathbf{r}_i$ with $n = j - i$. $n$ takes values from 0 to $m$-2. The square radius of gyration is defined as:

$$\langle R_g^2 \rangle = \frac{1}{m^2} \cdot \sum_{i<j}^{m} \langle \mathbf{r}_{i,j}^2 \rangle \tag{37}$$

where $\mathbf{r}_{i,j}$ is the vector joining monomer *i* and *j*. Depending on the model considered we have for ideal chains:

**FJC**

$$\langle \cos\theta(s) \rangle^* = 0 \tag{38}$$

$$\langle \mathbf{R}_e^2 \rangle^* = (m-1) \cdot l_b^2 = L \cdot l_b \tag{39}$$

$$\langle R_g^2 \rangle^* = \frac{1}{6} \cdot (m-1) \cdot \frac{m+1}{m} \cdot l_b^2 = \frac{1}{6} \cdot \frac{m+1}{m} \cdot L \cdot l_b \tag{40}$$

**FRC**

$$\langle \cos\theta(n) \rangle^* = (\cos\delta)^n \tag{41}$$

$$\frac{\langle \mathbf{R}_e^2 \rangle^*}{(m-1) \cdot l_b^2} = \frac{1+\cos\delta}{1-\cos\delta} - \frac{2 \cdot \cos\delta}{m-1} \cdot \frac{1-(\cos\delta)^{m-1}}{(1-\cos\delta)^2} \tag{42}$$

$$\frac{6 \cdot \langle R_g^2 \rangle^*}{(m-1) \cdot l_b^2} = \frac{m+1}{m} \cdot \frac{1+\cos\delta}{1-\cos\delta} - \frac{6}{m} \cdot \frac{\cos\delta}{(1-\cos\delta)^2} + \frac{12}{m^2} \cdot \frac{(\cos\delta)^2}{(1-\cos\delta)^3}$$
$$- \frac{12 \cdot (\cos\delta)^3}{(m-1) \cdot m^2} \cdot \frac{1-(\cos\delta)^{m-1}}{(1-\cos\delta)^4} \tag{43}$$

Equation (41) can be rearranged to introduce the persistent length of the FRC chain, $l_p$:

$$\langle \cos\theta(n) \rangle^* = \exp\left(-\frac{n \cdot l_b}{l_p}\right) \tag{44}$$

with

$$l_p = -\frac{l_b}{\ln(\cos\delta)} \tag{45}$$

In the limit of a WLC ($l_p \gg l_b$) equation (42) and (43) become:

$$\frac{\langle \mathbf{R}_e^2 \rangle^*}{2 \cdot l_p^2} = X - 1 + \exp(-X) \tag{46}$$

$$\frac{\langle R_g^2 \rangle^*}{2 \cdot l_p^2} = \frac{X}{6} - \frac{1}{2} + \frac{1}{X} - \frac{1}{X^2} \cdot (1 - \exp(-X)) \tag{47}$$

with $X = L/l_p$. For $X \ll 1$ a WLC appears rigid while for $X \gg 1$ it is viewed as flexible. In both cases it must be regarded as having no thickness.

Again star exponents denote ideal quantities.